\documentclass[pdflatex,sn-aps]{sn-jnl}

\usepackage{graphicx}%
\usepackage{multirow}%
\usepackage{amsmath,amssymb,amsfonts}%
\usepackage{amsthm}%
\usepackage{mathrsfs}%
\usepackage[title]{appendix}%
\usepackage{xcolor}%
\usepackage{textcomp}%
\usepackage{manyfoot}%
\usepackage{booktabs}%
\usepackage{listings}%
\usepackage{hyperref}
\usepackage{geometry}
\theoremstyle{thmstyleone}%
\theoremstyle{thmstyletwo}%

\theoremstyle{thmstylethree}%

\begin{document}

\title[Article Title]{\textbf{Analysis of the cosmological evolution parameters, energy conditions, and linear matter perturbations of an exponential-type model in $f(Q)$ gravity}}


\author*[1]{\fnm{Ivan R.} \sur{Vasquez}}\email{itsivanvasquez@gmail.com}

\author[1]{\fnm{A.} \sur{Oliveros}}\email{alexanderoliveros@mail.uniatlantico.edu.co}

\affil*[1]{\orgdiv{Grupo de Física de Partículas Elementales y Cosmología, Programa de Física}, \orgname{Universidad del Atlántico}, \orgaddress{\street{Carrera 30 No. 8-49}, \city{ Puerto Colombia}, \state{Atlántico}, \country{Colombia}}}


\abstract{We analytically study cosmological evolution in a flat FLRW spacetime in the context of modified STEGR gravity or $f(Q)$, using an exponential two-parameter model which represents a smooth perturbative expansion around the $\Lambda$CDM model. 
The cosmological analysis is carried out by calculating the Hubble parameter as a function of redshift, for selected values of the parameters. The Hubble parameter is obtained analytically by means of several approximations good enough to deviate slightly from the numerical solution.  Several late-time cosmological parameters are computed, such as dark energy state parameter, deceleration parameter, and statefinder parameters. Additionally, we analyzed the behavior of the classical energy conditions WEC, SEC, NEC, and DEC for both the combination of matter and geometrical contribution and the geometrical contribution alone. Beyond the background level, linear matter perturbations are studied by calculating parameters relevant to structure growth and formation. The overall results indicate that the model may exhibit quintessence-like and phantom-like behavior and it also impacts the growth of structures in the universe by means late-time contributions to clustering.}


\keywords{Cosmology, Modified gravity, Exponential $f(Q)$ gravity, Late-time acceleration}



\maketitle

\section{Introduction}\label{sec1}

The Universe appears to be accelerating its expansion, according to the celebrated observations in 1998 using Type IA Supernovae \cite{supernova1, supernova2}. Solely by this observation, a huge problem was created in modern cosmology, because nothing in the contents of the universe has the effect of accelerating the expansion, a problem which can be dubbed as: \textbf{the dark energy problem}. Here dark energy refers to a mysterious field or fluid underlying the cosmological dynamics at the present epoch (for a comprehensive review, see \cite{copeland, Bamba, peebles}). The current paradigm states that the universe is greatly dominated by a fluid of constant density and pressure coming from the cosmological constant term $\Lambda$ introduced by Einstein \cite{Einstein1, Einstein2}. Therefore, the dynamics of the universe is consistent with the introduction of the cosmological constant, constituting the so-called $\Lambda$CDM model \cite{lambdaterm}. On the other hand, in 2024, the DESI collaboration provided hints that dark energy may be dynamical and not the cosmological constant, as it is currently accepted \cite{desi}. 

In essence, the problem is to explain what kind of dynamics is causing such an effect, either by modifying the energy-momentum tensor $T_{\mu\nu}$; via including additional fields to source the mysterious dark energy (see \cite{copeland, Bamba} and references therein), or by modifying the Einstein-tensor $G_{\mu\nu}$, mainly by modifying the Einstein-Hilbert action. In this case, theories such as $f(R)$ \cite{defelice, Capozziello1}, as many other examples that fall in the case called \textit{modified gravity} \cite{Clifton}. Thus, in this effort to understand the dynamics of such a perplexing problem, many new alternative theories have arisen in the context of modified gravity. 

We are particularly focused on a new class of theories that appear as a subset of the so-called \textit{geometrical trinity of gravity}. In this framework, a new set of geometrical quantities arises and permits the construction of theories equivalent to Einstein's General Relativity, where gravity is described by these geometrical quantities rather than curvature (see these works for a complete review on the subject \cite{fqreview, trinity}). Therefore, extensions similar to $f(R)$ appear naturally, and we are left with extended theories such as: $f(T)$ (Extended Teleparallel Gravity), $f(Q)$ (Extended Symmetric Teleparallel Gravity), and combinations of those two. Here $T$ is a scalar called \textit{torsion} and $Q$ is called nonmetricity, and they appear as geometrical quantities in metric affine spaces \cite{coincident}. In this sense, we are particularly interested in $f(Q)$ theories as they have become regular, which since its introduction many models have appeared. A varied set of models can be found to correspond with GR but not $\Lambda$CDM; this in order to alleviate the use of the cosmological constant \cite{Myrzakulov, MandalQ, anagnos}. Usually, these models are of power-law type which is generally an approximation in exponential-type models \cite{anagnos}. In \cite{Mandal}, several models are presented, to which they compute $H(z)$ and analyze associated energy conditions, also in \cite{anagnos2} various models including an exponential model are considered to produce constraints using Big Bang Nucleosynthesis formalism. Another interesting model is a for which the background spacetime gives rise to the same Friedmann equations of GR. However, while studying perturbations in background FLRW spacetime several \textit{signatures} of $f(Q)$ were found in the cosmological observables \cite{signatures, Barros}. Unified dark-energy inflation models also exist, and extensions including Gauss-Bonnet invariant terms \cite{Odintsov1, Odintsov2}.

It is usually pertinent to study both background dynamics and perturbations in every model introduced. A recurrent practice is to obtain $H(z)$ analytically and numerically if necessary, and also from $H(z)$ datasets, a constraint on the parameters of the model is obtained. For example, in \cite{MandalQ}, authors found an analytical solution for $H(z)$ by using a parametrization of the state parameter $w_{DE}$. They proceeded to constrain the parameters of the model through Markov Chain Monte Carlo (MCMC) statistical analysis and use such constraint to explore dynamics in cosmological parameters. This technique is also used in the exponential-type model introduced in \cite{OliverosAcero}, which also explores several cosmological parameters to determine the viability of the proposed model, forming the base for this paper. 

In this work we further explore the model introduced by Oliveros and Acero \cite{OliverosAcero}, which is perturbative around $\Lambda$CDM and it allows analytical solutions for relevant cosmological parameters. We analytically extend their study as follows: In section \ref{stegr} we give a brief introduction to the non-metric theories and extended $f(Q)$ along its application to cosmology. In section \ref{modelexp} we propose a reasonable expansion of the $f(Q)$ functional in order to obtain analytical solutions for $H(z)$ for the parameter sets; $(n=1, b<1)$ and $(n=2, b<1)$, these solutions are tested against numerical solutions obtained by using the exact Friedmann differential equation. In section \ref{energycond}, we analyze the behavior of classical energy conditions as formulated in classical gravitation. Namely, WEC, SEC, NEC and DEC\footnote{WEC: Weak Energy Condition, SEC: Strong Energy Condition, NEC: Null Energy Condition, DEC: Dominant Energy Condition.}. We complement this part by including a calculation of the state parameter $w_{DE}$. Additionally, we include a subsection devoted to use Raychaudhuri's equation as a tool to confirm whether the model produces a late-time regime of accelerated expansion. Further, in section \ref{statefinder}, we explore statefinder parameters such as the set ($q$, $r$, $s$) which are computed and studied to determine their predictions against $\Lambda$CDM, also a statefinder diagnostic is performed in the $r-s$ plane providing the trajectories followed by the model during the cosmological evolution. Besides, we include another useful diagnostic called $Om(z)$ which helps to further characterize the equivalent dark energy model by analyzing the evolution of this parameter. In section \ref{perturbations}, we generally go over linear perturbations in the matter sector by numerically computing of fundamental observables with an impact on large scale structures. We compute the linear density fluctuations for matter numerically from the evolution differential equation, this equation includes an effective gravitational constant affected by the form of $f(Q)$ whose evolution is studied and its impact on structure formation is analyzed. A proposed functional for $f_{g}$ allows to obtain the growth-index parameter which is $\gamma$. To complement our study, we compute and analyze the $f\sigma_{8}$ parameter and compare it to observational data. Finally, in section \ref{conclusions} we present a discussion and overall conclusions about our work.

\section{STEGR and $f(Q)$ extensions}\label{stegr}
\noindent In general, the action for an $f(Q)$ gravity model in the presence of matter:
\begin{equation}\label{eq_action}
S = \int_{\mathcal{M}}{d^4x \sqrt{-g} \left(\frac{f(Q)}{16\pi G} + \mathcal{L}_{\rm{M}}\right)},
\end{equation}
where $g$ denotes the determinant of the metric tensor, $g^{\mu\nu}$, with $G$ being the Newton's constant (from now on, we will use geometrized units $G=c=1$), unless stated otherwise. $\mathcal{L}_{\rm{M}}$ represents the Lagrangian density for the matter components (relativistic and non-relativistic perfect matter fluids). The term $f(Q)$ is for now an arbitrary functional of the non-metricity scalar $Q$, and the integral is performed over all spacetime $\mathcal{M}$. This scalar is obtained by the contraction of the non-metricity tensor, which is defined as \cite{fqreview, trinity, coincident}
\begin{equation}\label{eq_Qdef}
Q_{\gamma\mu\nu} \equiv \nabla_\gamma g_{\mu\nu} = \partial_\gamma g_{\mu\nu} - \Gamma^{\beta}_{\mkern10mu\gamma\mu}g_{\beta\nu} - \Gamma^{\beta}_{\mkern10mu\gamma\nu}g_{\beta\mu},
\end{equation}
where $\nabla_\gamma$ represents the covariant derivative with respect to a general affine connection $\Gamma^{\alpha}_{\mkern10mu\gamma\beta}$, which can be written as \cite{hehl, ortin}
\begin{equation}\label{eq_affine}
\Gamma^\lambda\\
_{\mu\nu}=\left\{^\lambda\\
_{\mu\nu}\right\}+K^\lambda\\
_{\mu\nu}+L^\lambda\\
_{\mu\nu},
\end{equation}
with the Levi-Civita connection of the metric given by
\begin{equation}\label{eq_LeviCivita}
\left\{^\lambda\\
_{\mu\nu}\right\}\equiv\frac{1}{2}g^{\lambda\beta}\left(\partial_\mu g_{\beta\nu}+\partial_\nu g_{\beta\mu}-\partial_\beta g_{\mu\nu}\right);
\end{equation}
the contortion tensor is
\begin{equation}\label{eq_contortion}
K^\lambda\\
_{\mu\nu}\equiv\frac{1}{2}T^\lambda\\
_{\mu\nu}+T^{\,\,\,\,\,\lambda}_{(\mu\,\,\,\,\nu)},
\end{equation}
with the torsion tensor $T^\lambda_{\,\,\,\,\mu\nu} \equiv 2\Gamma^\lambda_{\,\,\,\,[\mu\nu]}$, and the disformation tensor $L^\lambda_{\,\,\,\,\mu\nu}$, which can be written in terms of the non-metricity tensor as
\begin{equation}\label{eq4}
L^{\beta}_{\mkern10mu\mu\nu}=\frac{1}{2}Q^{\beta}_{\mkern10mu\mu\nu}-Q_{(\mu\nu)}^{\mkern35mu\beta}.
\end{equation}
The contraction of the tensor $Q_{\gamma\mu\nu}$ is given by
\begin{equation}\label{eq2}
Q = -Q_{\gamma\mu\nu} P^{\gamma\mu\nu},
\end{equation}
where the tensor $P^{\gamma\mu\nu}$ is the non-metricity conjugate, defined as
\begin{equation}\label{eq_metricityConj}
P^{\beta}_{\mkern10mu\mu\nu}=-\frac{1}{2}L^{\beta}_{\mkern10mu\mu\nu}+\frac{1}{4}(Q^\beta-\tilde{Q}^\beta)g_{\mu\nu}
-\frac{1}{4}\delta^{\beta}_{\,\,\,(\mu}Q_{\nu)},
\end{equation}
the other quantities shown in Eq.~(\ref{eq_metricityConj}) are the contractions of non-metricity with the metric tensor, given by:
\begin{equation}\label{eq5}
Q_{\beta}=g^{\mu\nu}Q_{\beta\mu\nu},\quad \tilde{Q}_{\beta}=g^{\mu\nu}Q_{\mu\beta\nu}.
\end{equation}
Variation of the action (\ref{eq_action}) with respect to the metric gives the equation of motion \cite{trinity, coincident},
\begin{equation}\label{eq_MotionEq}
\frac{2}{\sqrt{-g}}\nabla_\beta(f_Q\sqrt{-g}P^\beta_{\mkern10mu\mu\nu}) + \frac{1}{2}fg_{\mu\nu} + f_Q(P_{\mu\beta\lambda}Q_\nu^{\mkern10mu\beta\lambda}-2Q_{\beta\lambda\mu}P^{\beta\lambda}\mkern1mu\nu) = -T_{\mu\nu},
\end{equation}
where $f_Q \equiv \frac{df}{dQ}$, and as usual, the energy momentum tensor $T_{\mu\nu}$ is given by
\begin{equation}\label{eq_Tmunu}
T_{\mu\nu}=-\frac{2}{\sqrt{-g}}\frac{\delta\sqrt{-g}\mathcal{L}_{\rm{M}}}{\delta g^{\mu\nu}},
\end{equation}
which in the cosmological context is considered to be a perfect fluid, i.e.,~$T_{\mu\nu}=(\rho+p)u_\mu u_\nu+pg_{\mu\nu}$, where $p$ and $\rho$ are the total pressure and total energy density of any perfect fluid of matter, radiation and DE, respectively. The quantity $u^\mu$ represents the 4-velocity of the fluid.

Considering the flat Friedman-Robertson-Walker (FRW) metric,
\begin{equation}\label{eq_FRWmetric}
ds^2 = -dt^2 + a^2(t)\delta_{ij}dx^idx^j,
\end{equation}
with $a(t)$ representing the scale factor, the time and spatial components of Eq. (\ref{eq_MotionEq}) are given, respectively, by \cite{fqreview, trinity}
\begin{equation}\label{eq_timeEq}
6f_QH^2 - \frac{1}{2}f = 8\pi(\rho_{\rm{m}} + \rho_{\rm{r}}),
\end{equation}
and 
\begin{equation}\label{eq_spaceEq}
(12H^2f_{QQ} + f_{Q})\dot{H} = -4\pi\left(\rho_{\rm{m}} + \frac{4}{3}\rho_{\rm{r}}\right),
\end{equation}
where $\rho_{\rm{m}}$ is the matter density and $\rho_{\rm{r}}$ denotes the density of radiation. $f_{QQ} = \frac{d^2f}{dQ^2}$, the over-dot denotes a derivative with respect to cosmic time $t$ and $H\equiv \dot{a}/a$ is the Hubble parameter. Further, the non-metricity scalar associated with the metric (\ref{eq_FRWmetric}) is given by
\begin{equation}\label{eq_Qscalar}
Q=6H^2.
\end{equation}
It should be noted that in this scenario, the standard Friedmann equations of General Relativity (GR) plus the cosmological constant are recovered when assuming that $f(Q)=Q+2\Lambda$.

If there is no interaction between non-relativistic matter and radiation, then these components obey separately the conservation laws
\begin{equation}\label{eq11}
\dot{\rho}_{\rm{m}} + 3H\rho_{\rm{m}} = 0,\quad \dot{\rho}_{\rm{r}} + 4H\rho_{\rm{r}} = 0.
\end{equation}
As usual in the literature, it is possible to rewrite the field equations (\ref{eq_timeEq}) and (\ref{eq_spaceEq}) in the Einstein-Hilbert form, respectively:
\begin{equation}\label{eq12}
3H^2 = 8\pi\rho,
\end{equation}
\begin{equation}\label{eq13}
\dot{H} = -4\pi(\rho + p),
\end{equation}
where $\rho = \rho_{\rm{m}}+ \rho_{\rm{r}} + \rho_{\rm{DE}}$ and $p = p_{\rm{m}} + p_{\rm{r}} + p_{\rm{DE}}$ correspond to the total effective energy pressure densities of the cosmological fluid, respectively.  In this case, the dark energy component has a geometric origin, and after some manipulation in Eqs.~(\ref{eq_timeEq})  and (\ref{eq_spaceEq}), we obtain the effective dark energy and pressure corresponding to the $f(Q)$-theory given by
\begin{equation}\label{eq_effRhoDE}
\rho_{\rm{DE}} = \frac{1}{8\pi}\left[\frac{1}{2}f + 3H^2(1-2f_Q)\right],
\end{equation}
and
\begin{equation}\label{eq_effPDE}
p_{\rm{DE}} = \frac{1}{4\pi}\left[2\dot{H}(12H^2f_{QQ} + f_Q - 1) - \rho_{\rm{DE}}\right].
\end{equation}
It is easy to show that $\rho_{\rm{DE}}$ and $p_{\rm{DE}}$ defined in this way satisfy the usual energy conservation equation
\begin{equation}\label{eq16}
\dot{\rho}_{\rm{DE}}+3H(\rho_{\rm{DE}}+p_{\rm{DE}})=0.
\end{equation}
In this case, we assume that the equation of state parameter for this effective dark energy satisfies the following relation:
\begin{equation}\label{eq17}
w_{\rm{DE}} = \frac{p_{\rm{DE}}}{\rho_{\rm{DE}}} = -1 + \frac{2\dot{H}(12H^2f_{QQ} + f_Q - 1)}{\frac{1}{2}f + 3H^2(1 - 2f_Q)}.
\end{equation}
Here we recover $w_{\rm{DE}}=-1$ ($\Lambda$CDM) for $f_{Q}=1$ and $f_{QQ}=0$.

\section{Exponential-type $f(Q)$ model in flat FLRW metric}\label{modelexp}
From modified Friedmann equations in the $f(Q)$ scenario, we consider a cosmological model with an exponential-type, well-behaved function as follows:
\begin{align}
    f(Q)=Q+Ae^{g(Q)},
\end{align}
where $A$ is a constant and $g(Q)$ a smooth function of the non-metricity scalar $Q$. Corresponding to the $\Lambda$CDM model, we ask that for $g(Q)\rightarrow 0$, then $f(Q)\rightarrow Q+2\Lambda$. Therefore we fix, $A=2\Lambda$ as in GR/STEGR. Models of the exponential type seem to be favoured by latest Pantheon+ data and DESI (see Odintsov, \textit{et al.} in [arXiv:2412.09409 [gr-qc]]). Before specializing to a particular form of $g(Q)$, we obtain Friedmann equations for an exponential-type model, given by:
\begin{align}
     \label{fried1}
    3H^2+\Lambda e^{g}(12H^2g'-1)=8\pi \rho, \\
    \label{fried2}
    24H^2\Lambda e^{g}[g''+(g')^2]\dot{H}+\dot{H}+2\Lambda e^{g}g'\dot{H}=-4\pi(\rho+p).
\end{align}
Here $g'=d_{Q}g, g''=d^2_{Q}g$. From these two equations we can obtain both \textit{dark energy} density and its corresponding pressure,
\begin{align}
\label{energydensity}
    \rho_{DE}=\frac{\Lambda}{8\pi}e^{g}(1-12H^2g'),\\
    p_{DE}=\frac{\Lambda}{8\pi}e^{g}\left(48\dot{H}[g''+g'^2]+4g'\dot{H}+36H^2g'-1\right).
\end{align}
We are assigning energy density and pressure for the contributions of modified gravity in this context. The state parameter $w_{DE}$ results in:
\begin{align}
w_{DE}=\frac{48\dot{H}[g''+g'^2]+4g'(\dot{H}+9H^2)-1}{(1-12H^2g')}.
\end{align}

For smoothness, $g(Q)$ must not diverge even if $Q$ diverges, therefore we can consider that as $Q\rightarrow \infty$ then $g(Q)\rightarrow 0$. This condition ensures that $\Lambda$ dominates at late cosmological times, resulting in the known state parameter $w_{DE}=-1$ in the $\Lambda$CDM model. The form of the function $g(Q)$ in this case should be an inverse power-type law, $g(Q)\sim Q^{-n}$. 

A nice model to be worked out has been introduced recently in the literature by Oliveros and Acero \cite{OliverosAcero}, heavily inspired by other works in $f(R)$ gravity \cite{Granda, OliverosAcero2, Oliveros1, Cognola, Odintsov}. The model fulfills the conditions discussed above, and it is given by:
\begin{align}
    f(Q)=Q+2\Lambda e^{-(b\Lambda/Q)^n},
\end{align}
which forms two-parameter model with $b,n$ as real parameters; here $n>0$ while $b\neq 0$, being a deviation parameter from $\Lambda$CDM. This model is basically a perturbative expansion around the $\Lambda$CDM Lagrangian, as a power series expansion reveals:
\begin{align}
    f(Q)=Q+2\Lambda \left(1-\frac{k^n}{Q^n}+\frac{1}{2!}\frac{k^{2n}}{Q^{2n}}+...\right); \quad k=b\Lambda.
\end{align}
Statistically, this model fits well with data for 
$H(z)$ generally preferring negative values of parameter $b$ according to the study carried out using Markov Chain Monte Carlo method, which also serves as a basis for this work \cite{OliverosAcero}. Other similar exponential models also seem to be statistically favoured by data according to the study carried out in \cite{Khyllepexpo}. Inserting the model in (\ref{fried1}) we obtain the following transcendental equation for $H(t)$ and recalling that $Q=6H^2$:
\begin{align}
    3H^2+\Lambda e^{-(b\Lambda/6H^2)^{n}}\left[12H^2n(b\Lambda)^{n}(6H^2)^{-n-1}-1\right]=8\pi \rho,
\end{align}
in order to obtain an algebraic equation for $H(t)$ it is natural to consider approximations by constraining the set of parameters $(b,n)$. In particular, let us consider that $b\Lambda/6H^2<<1$, and also rewrite the equation in terms of cosmological density parameters $\Lambda=3H^2_{0}\Omega_{\Lambda,0}$ and $\Omega_{\Lambda,0}=1-\Omega_{m,0}$ (by neglecting contributions of radiation $\Omega_{r,0}$). For generalizing purposes, we can thus expand the exponential up to second order,
\begin{align}
    e^{-(b\Lambda/6H^2)^n}\approx 1-\left(\frac{H^2_{0}\Omega_{\Lambda,0}b}{2H^2}\right)^n+\frac{1}{2}\left(\frac{H^2_{0}\Omega_{\Lambda,0}b}{2H^2}\right)^{2n}.
\end{align}
The condition that $bH^2_{0}\Omega_{\Lambda,0}/2H^2 \ll 1$ implies that if we consider small values of the parameter $b$, then at late times, when $H^2 \sim H^2_{0}\Omega_{\Lambda,0}$, the condition is satisfied. Similarly, during the matter-dominated era, where $H^2 \gg H^2_{0}\Omega_{\Lambda,0}$, the condition also holds. Finally, since it is customary in cosmological observations, the parameter to measure evolution is the redshift $z$ then the Hubble parameter is a function $H=H(z)$. We thus obtain the equation,
\begin{multline}\label{master}
3H^2+3H^2_{0}\Omega_{\Lambda}\left(1-\left(\frac{H^2_{0}\Omega_{\Lambda}b}{2H^2}\right)^n+\frac{1}{2}\left(\frac{H^2_{0}\Omega_{\Lambda}b}{2H^2}\right)^{2n}\right)[12H^2 n (H^2_{0}\Omega_{\Lambda}b)^{n}(2H^2)^{-n-1}-1]\\
\approx 3H^2_{0}\Omega_{m,0}(1+z)^3.
\end{multline}
From this equation, we should be able to obtain $H(z)$ analytically for certain approximations and values of the parameter set $(b,n)$. In this sense, our approach is similar to \cite{OliverosAcero} by obtaining analytical algebraic solutions in the case of $n=1$, then also extending the analytical solutions obtained including $n=2$. It should be noted that for odd values of $n$, if the parameter $b$ is negative of the form $b=-\lambda^2$ and $n=2k+1$, then it is clear that  
$e^{H^2_{0}\Omega_{\Lambda,0}\lambda^2/2H^2}e^{-(-H^2_{0}\Omega_{\Lambda,0}\lambda^2/2H^2)^{2k}} > e^{-H^2_{0}\Omega_{\Lambda,0}\delta^2/2H^2}e^{-(H^2_{0}\Omega_{\Lambda,0}\delta^2/2H^2)^{2k}}$,  
where for positive values of $b$, we adopt the form $b=\delta^2$. During the matter-dominated era, the deviation from the $\Lambda$CDM model is very small, while at late times, the effect of the deviation is greater for negative values of $b$. This is because if $\lambda=\delta$, then at late times, we have $e^{\lambda^2/2} > e^{-\lambda^2/2}$.

\section{Analytical and numerical solutions for $H(z)$}
In this section we aim to obtain analytical solutions for $H(z)$, analyze its behaviour and use it to perform other cosmological calculations. To proceed, we first consider the case $n=1, b<1$, which by the form of the Lagrangian the sign of $b$ should affect the encountered dynamics.
\subsection{Case I: $n=1, b<1$}
Equation (\ref{master}) for the case that $n=1$, can be written as follows:
\begin{multline}
3H^2+3H^2_{0}\Omega_{\Lambda,0}\left(1-\left(\frac{H^2_{0}\Omega_{\Lambda,0}b}{2H^2}\right)+\frac{1}{2}\left(\frac{H^2_{0}\Omega_{\Lambda,0}b}{2H^2}\right)^{2}\right)\left[\frac{H^2_{0}\Omega_{\Lambda,0}b}{H^2}-1\right]\\
\approx 3H^2_{0}\Omega_{m,0}(1+z)^3.
\end{multline}
We first consider $b<1$ and keep values up to second order in $b$. Therefore we get,
\begin{align}
3H^2+3H^2_{0}\Omega_{\Lambda,0}\left(1-\left(\frac{H^2_{0}\Omega_{\Lambda,0}b}{2H^2}\right)\right)\left[\frac{H^2_{0}\Omega_{\Lambda,0}b}{H^2}-1\right]\approx 3H^2_{0}\Omega_{m,0}(1+z)^3,
\end{align}
where simple algebra and the change of variable $u(z;b)=H^2/H^2_{0}$ leaves us with the following,
\begin{align}
    u-\xi(z)+\frac{3}{2}u^{-1}\Omega^2_{\Lambda,0}b-\frac{5}{8}u^{-2}\Omega^{3}_{\Lambda,0}b^2=0,
\end{align}
where $\xi(z)=\Omega_{m,0}(1+z)^3+\Omega_{\Lambda,0}$ and this simply results in a cubic equation for $u(z;b)$ by multiplying for $u^2(z;b)$ through the whole equation. And finally,
\begin{align}\label{cubic}
    u^3-\xi(z)u^2+\frac{3}{2}u\Omega^2_{\Lambda,0}b-\frac{5}{8}\Omega^{3}_{\Lambda,0}b^2=0.
\end{align}
This equation can be solved by employing a similar method used in \cite{OliverosAcero}, introduced in \cite{Basilakos} for $f(R)$. In our case, we propose for $u(z;b)$ an expansion around $u(z;0)=u_{\Lambda}=\xi(z)$ (corresponding to the $\Lambda$CDM model), of the form:
\begin{align}
    u(z;b)=u_{\Lambda}+\sum^{n}_{i=1}b^{i}\delta u_{i}(z;b),
\end{align}
where our expansion goes up to second order in $b$. Plugging this back into (\ref{cubic}), we separate the orders of expansion as,
\begin{align}
    \textbf{0th:} \quad u^3_{\Lambda}-\xi(z)u^2_{\Lambda}=0 \quad (\Lambda\rm{CDM}), \\
    \textbf{1st:} \, 3u^2_{\Lambda}\delta u_{1}-2\xi(z)u_{\Lambda}\delta u_{1}+\frac{3}{2}u_{\Lambda}\Omega^2_{\Lambda,0}=0, \\
    \textbf{2nd:} \, 3u^{2}_{\Lambda}\delta u_{2}+3(\delta u_{1})^2u_{\Lambda}-2\xi(z)u_{\Lambda}\delta u_{2}-\xi(z)(\delta u_{1})^2+\frac{3}{2}\delta u_{1}\Omega^2_{\Lambda,0}=\frac{5}{8}\Omega^3_{\Lambda,0}.
    \end{align}
The first order term clearly corresponds to the $\Lambda$CDM case (unperturbed case), while at first and second order in $b$ we obtain $\delta u_{1}, \delta u_{2}$ as:
\begin{align}
    \delta u_{1}=-\frac{7\Omega^2_{\Lambda,0}}{2\xi(z)},\\
    \delta u_{2}=\frac{\Omega^3_{\Lambda,0}(5\xi(z)-18\Omega_{\Lambda,0})}{8\xi^3(z)}.
\end{align}
Therefore the final solution up to second order reads, for $H^2(z)$ as:
\begin{align}\label{firstorder}
    H^2(z;b)=H^2_{\Lambda}+\frac{3H^2_{0}b\Omega^2_{\Lambda,0}}{2\xi(z)}\left[\frac{\Omega_{\Lambda,0}(5\xi(z)-18\Omega_{\Lambda,0})}{12\xi^2(z)}b-1\right].
\end{align}
From which one finds that for $b=0$, then $H^2(z)=H^2_{0}\xi(z)$, recovering the $\Lambda$CDM result. For the present value, $H^2(z=0;b)$ depends on the perturbative parameter $b$ as follows:
\begin{align}\label{initialhubble}
    H^2_{0}(b)=H^{2}_{0}\left[1-\frac{3\Omega^2_{\Lambda,0}}{2}b+\frac{\Omega^3_{\Lambda,0
    }(5-18\Omega_{\Lambda,0})}{8}b^2\right],
\end{align}
in this case the value of Hubble's constant depends on the sign of parameter $b$, in this way it could be greater than or lower than $H_{0}$. In another interesting limit we can consider high redshift regimes, for which $\xi(z)\rightarrow \infty$ and $H^{2}(z;b)\rightarrow H^2_{\Lambda}(z)$, behaving as in $\Lambda$CDM. In the de Sitter limit with $\xi(z)\rightarrow \Omega_{\Lambda}=1$, therefore at late times we have,
\begin{align}
    H^2_{dS}=H^2_{0}\left(1-\frac{3}{2}b-\frac{13}{8}b^2\right).
\end{align}
From previous expressions we can obtain a two overlapping constraints for the parameter $b$. That is, with the condition that $H^2_{0}(b)>0$ we obtain $b\in (-3.391, 1.001)$ and with $H^2_{dS}>0$ we get, $b\in (-1.371, 0.448)$. If we refer to the constraints obtained in \cite{OliverosAcero}, the best-fit values from the MCMC analysis fall into these analytical constraints. The exact algebraic solutions of the cubic equation (\ref{cubic}) result in a solution given by the following expression,
\begin{multline}
u(z;b)=\frac{\xi(z)}{3} - \frac{2^{1/3} \left( \frac{9 b \Omega_{\Lambda,0}^2}{2} - \xi^2 \right)}
{3 \left( \frac{135 b^2 \Omega_{\Lambda,0}^3}{8} - \frac{27}{2} b \Omega_{\Lambda,0}^2 \xi + 2 \xi^3 + 
\sqrt{4 \left( \frac{9 b \Omega_{\Lambda,0}^2}{2} - \xi^2 \right)^3 + 
\left( \frac{135 b^2 \Omega_{\Lambda,0}^3}{8} - \frac{27}{2} b \Omega_{\Lambda,0}^2 \xi + 2 \xi^3 \right)^2} \right)^{1/3}} \\
+ \frac{\left( \frac{135 b^2 \Omega_{\Lambda,0}^3}{8} - \frac{27}{2} b \Omega_{\Lambda,0}^2 \xi + 2 \xi^3 + \sqrt{4 \left( \frac{9 b \Omega_{\Lambda,0}^2}{2} - \xi^2 \right)^3 + 
\left( \frac{135 b^2 \Omega_{\Lambda,0}^3}{8} - \frac{27}{2} b \Omega_{\Lambda,0}^2 \xi + 2 \xi^3 \right)^2} \right)^{1/3}}{3 \cdot 2^{1/3}},
\end{multline}
such that $u(z;b=0)$ reduces to $u(z;0)=\xi(z)$ in correspondence with the $\Lambda$CDM model. The other two solutions are complex and have the form,
\begin{multline}
\frac{\xi}{3} + \frac{(1 \pm i \sqrt{3}) \left( \frac{9 b \Omega_{\Lambda,0}^2}{2} - \xi^2 \right)}
{3 \cdot 2^{2/3} \left( \frac{135 b^2 \Omega_{\Lambda,0}^3}{8} - \frac{27}{2} b \Omega_{\Lambda,0}^2 \xi + 2 \xi^3 + 
\sqrt{4 \left( \frac{9 b \Omega_{\Lambda,0}^2}{2} - \xi^2 \right)^3 + 
\left( \frac{135 b^2 \Omega_{\Lambda,0}^3}{8} - \frac{27}{2} b \Omega_{\Lambda,0}^2 \xi + 2 \xi^3 \right)^2} \right)^{1/3}} \\
- \frac{(1 \mp i \sqrt{3}) \left( \frac{135 b^2 \Omega_{\Lambda,0}^3}{8} - \frac{27}{2} b \Omega_{\Lambda,0}^2 \xi + 2 \xi^3 + 
\sqrt{4 \left( \frac{9 b \Omega_{\Lambda,0}^2}{2} - \xi^2 \right)^3 + 
\left( \frac{135 b^2 \Omega_{\Lambda,0}^3}{8} - \frac{27}{2} b \Omega_{\Lambda,0}^2 \xi + 2 \xi^3 \right)^2} \right)^{1/3}}{6 \cdot 2^{1/3}}.
\end{multline}
Thus, only one physical solution exists, which, when expanded in powers of $b$, results in the terms,
\begin{align}
u(z;b)=- \frac{3 b \Omega_{\Lambda,0}^2}{2 \xi} + \xi 
+ \frac{b^2 \Omega_{\Lambda,0}^3 (-18 \Omega_{\Lambda,0} + 5 \xi)}{8 \xi^3} 
+ \frac{9 b^3 \Omega_{\Lambda,0}^5 (-12 \Omega_{\Lambda,0} + 5 \xi)}{16 \xi^5}+...,
\end{align}
where considering only terms up to $b^2$ reduces to the analytical solution (\ref{firstorder}).

\subsection{Case II: $n=2, b<1$}
In this case, and similarly as previous subsection, equation (\ref{master}) up to second order in $b$ can be written as:
\begin{align}
    3H^2-3H^2_{0}\Omega_{\Lambda,0}\left(1-\left(\frac{H^2_{0}\Omega_{\Lambda,0}b}{2H^2}\right)^2\right)\left[\frac{H^4_{0}\Omega_{\Lambda,0}^2b^2}{H^4}-1\right]\approx 3H^2_{0}\Omega_{m,0}(1+z)^3,
\end{align}
where again we obtain an algebraic equation for $u(z)$, given by:
\begin{align}
    u^3(z)-u^2(z;b)\xi(z)+\frac{5}{4}\Omega^3_{\Lambda,0}b^2=0.
\end{align}
Following a similar procedure as for the $n=1$ solution, we obtain again the Hubble parameter, now with the form:
\begin{align}\label{secondorder}
    H^2(z;b)=H^{2}_{0}\left(\xi(z)-\frac{5\Omega^{3}_{\Lambda,0}b^2}{4\xi^{2}(z)}\right)
\end{align}
The crucial difference with the solution (\ref{firstorder}) is that the values of $H(z;b), H_{0}(b)$ are insensitive to the sign of $b$. One may also obtain a range of values for parameter $b$ using the de Sitter condition, $H^2_{dS}>0$, so that we find the range to be $b\in(-0.894, 0.894)$. Similarly to the case $n=1$, the equation obtained for $u(z;b)$ has three solutions, where the only real solution is given by,
\begin{multline}
u(z;b)=\frac{1}{6} \Bigg( 
    2 \xi 
    + \frac{4 \xi^2}
    {\left(-135 b^2 \Omega_{\Lambda,0}^3 + 8 \xi^3 
    + 3 \sqrt{15} \sqrt{135 b^4 \Omega_{\Lambda,0}^6 - 16 b^2 \Omega_{\Lambda,0}^3 \xi^3} \right)^{1/3}} \\
    + \left(-135 b^2 \Omega_{\Lambda,0}^3 + 8 \xi^3 
    + 3 \sqrt{15} \sqrt{135 b^4 \Omega_{\Lambda,0}^6 - 16 b^2 \Omega_{\Lambda,0}^3 \xi^3} \right)^{1/3} 
\Bigg).
\end{multline}
where an expansion around $b=0$ once again leads to the following terms,
\begin{align}
u(z;b)=- \frac{25 b^4 \Omega_{\Lambda,0}^6}{8 \xi^5} 
- \frac{5 b^2 \Omega_{\Lambda,0}^3}{4 \xi^2} 
+ \xi(z)+...,
\end{align}
which contains the solution obtained in (\ref{secondorder}).

\subsection{Numerical solutions for $H(z)$}
\begin{figure}[h!]
    \includegraphics[width=0.5\linewidth]{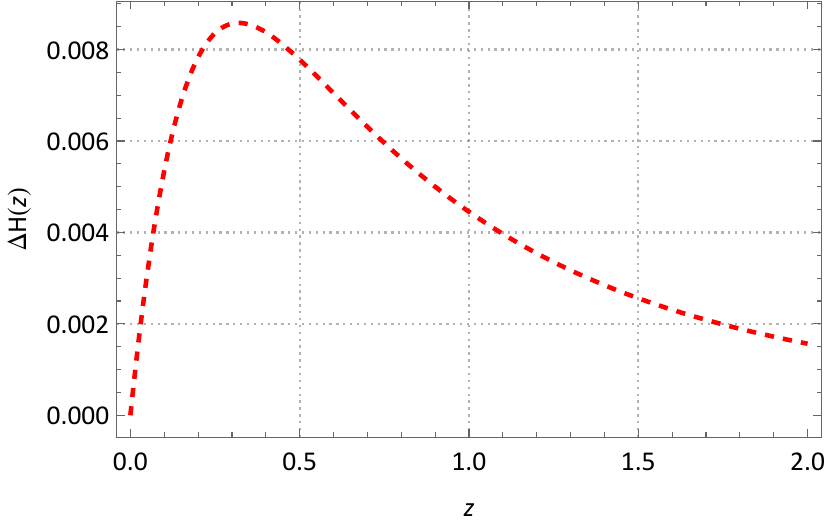}
    \hspace{0.1 cm}
    \includegraphics[width=0.5\linewidth]{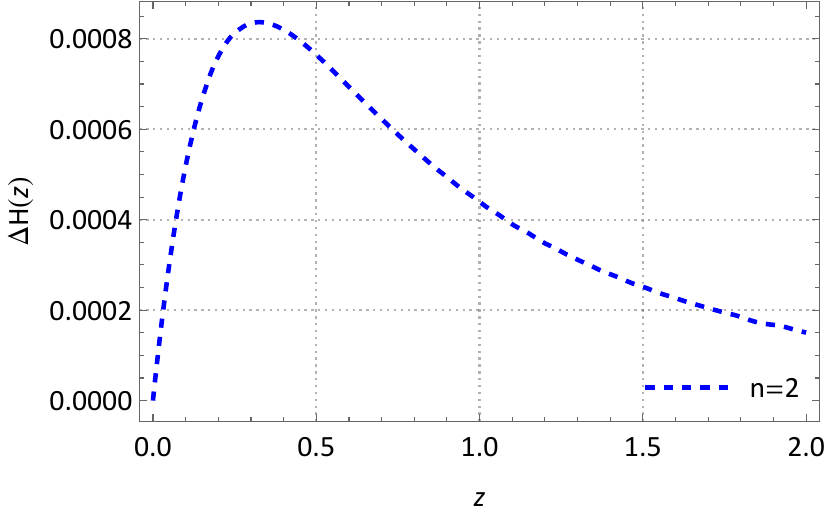}
    \caption{$\Delta H(z)$ vs. $z$ for the analytical solution with $b=0.1, n=2$ (Left). $\Delta H(z)$ vs. $z$ for the analytical solution with parameters $b=-0.1, n=1$ (Der).}
    \label{numerical}
\end{figure}

We have found analytical solutions for $H(z)$ in the Friedmann equations at first and second order. It is necessary to obtain a numerical solution of the full differential Friedmann equation for $H(z)$, and we proceed by considering $g(Q)=-(b\Lambda/Q)^{n}$ and calculate $g', g''$ and $(g')^2$:
\begin{align}
    g'=n(b\Lambda)^{n}Q^{-(n+1)}, \\
    g''=-n(n+1)(b\Lambda)^{n}Q^{-(n+2)}, \\
    g'^{2}=n^{2}(b\Lambda)^{2n}Q^{-2(n+1)}.
\end{align}
We insert back into (\ref{fried2}), recalling that $\dot{H}=-(1+z)H(z)H'(z)$ (where $H'(z)=d_{z}H(z)$) and since we are analyzing late times, we may neglect the contributions from radiation, and finally taking $n=1$. Therefore we obtain in exact form, a non-linear ordinary differential equation for $H(z)$ to be solved numerically,
\begin{multline}
\left[e^{-H^2_{0}\Omega_{\Lambda,0}b/2H^2}b\left(\frac{2H^4_{0}\Omega_{\Lambda,0}}{3H^4}+\frac{bH^6_{0}\Omega_{\Lambda,0}}{18H^{6}}\right)+1-\frac{b\Omega^2_{\Lambda,0}H^4_{0}}{2H^4}e^{-H^2_{0}\Omega_{\Lambda,0}b/2H^2}\right]H(z)H'=\\
\frac{3}{2}\Omega_{m,0}(1+z)^2.
\end{multline}
This is the differential Friedmann equation, and it gives $H(z)$, by supplying initial condition from equation (\ref{initialhubble}) $H_{0}(-0.1)=71.99 \,\rm{km}\rm{s}^{-1}\rm{Mpc}^{-1}$ and parameters, $\Omega_{m,0}=0.315, \Omega_{\Lambda,0}=0.685$ and $(n=1, b=-0.1)$. In order to compare the numerical solution with the previously obtained analytical solutions, we introduce the function $\Delta H(z)$, which shows the relative error as used in \cite{Oliveros1},
\begin{align}
    \Delta H(z)=\frac{100|H_{A}-H_{N}|}{H_{N}(z)},
\end{align}
where $H_{A}$ and $H_{N}$ are the analytical and numerical solutions respectively. We plot $\Delta H(z)$ and notice that the maximum relative error is at most $\sim 0.008 \%$ showing that the solution works well with the parameters used (see figure \ref{numerical}). In this way, we can use it for the analysis of other cosmological observables and by taking advantage of its analytical form we may obtain known limits for these cosmological parameters.

Similarly we obtain the numerical solution for $n=2$ with $b=0.1$ and the corresponding relative error with respect to analytical solution (\ref{secondorder}). As it is shown in figure \ref{numerical}, both solutions behave very similar to each other. In the case of the relative error for the $n=2$ analytical solution it has a maximum of about $\gtrsim 0.0008 \%$ working properly for the parameters used.
\section{Energy conditions in the $f(Q)$ model}\label{energycond}

Classical energy conditions in general relativity and extended modified theories of gravity alike, represent constraints on the matter-energy content in spacetime and help in characterizing the attractive nature of gravity \cite{poisson}. In other aspects, energy conditions help to define the geodesic structure of spacetime as they play a crucial role in singularity theorems. In cosmology, as it was originally formulated in Raychaudhuri's equation \cite{Raychaudhuri}, they help to discern between cosmological scenarios in certain violations of the conditions \cite{Santos}. It is also a tool for constraining modified gravity scenarios such as $f(R)$ and $f(Q)$ \cite{Mandal, Capozziello}. Considering the energy-momentum tensor $T^{(DE)}_{\mu\nu}$ of the geometrical contribution to \textbf{dark energy} (DE) as:
\begin{align}
    T^{(DE)}_{\mu\nu}=(\rho_{DE}+p_{DE})u_{\mu}u_{\nu}+p_{DE}g_{\mu\nu},
\end{align}
where $u_{\mu}$ is the tangent vector the timelike curve $\gamma$ of a co-moving observer in FLRW spacetime. Taking into account that $u_{\mu}u^{\mu}=-1$, we obtain the following energy conditions.

\subsection{Weak energy condition (WEC)}
\begin{figure}[h!]
\includegraphics[width=0.5\linewidth]{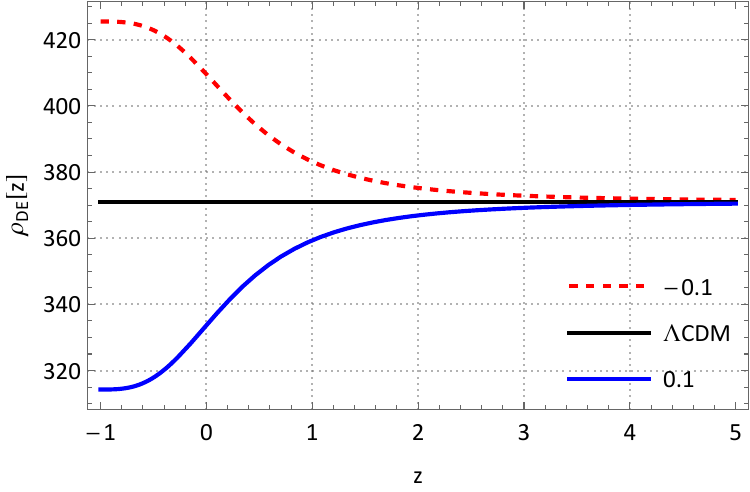} 
\hspace{0.1 cm}
\includegraphics[width=0.5\linewidth]{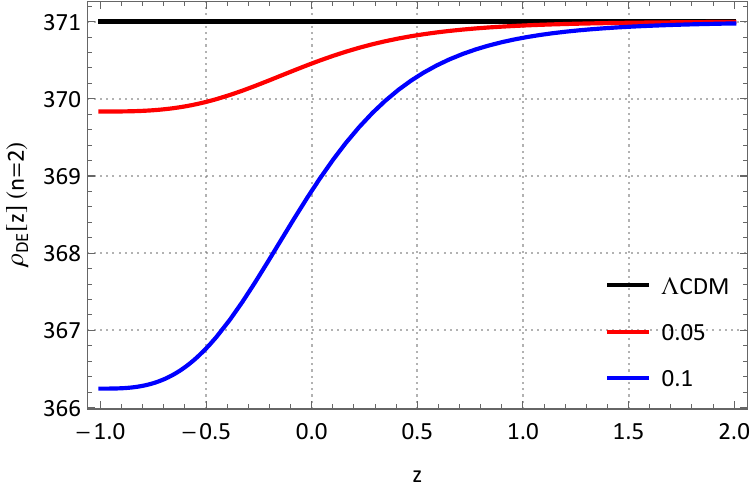}
\caption{Dark energy density $\rho_{DE}$ vs. $z$ for analytical solution with $n=1$, parametrized by $b$. We have used $\Omega_{m,0}=0.315$ and $H_{0}=67.36 km/s \operatorname{Mpc}^{-1}$ (Left). Dark energy density for analytical solution with $n=2$ (Der).}
\label{darkdensity}
\end{figure}

The weak energy condition or WEC states that the following inequality holds,
\begin{align}
    T^{(DE)}_{\mu\nu}u^{\mu}u^{\nu}\geq 0.
\end{align}
This becomes a statement about the positivity of energy-density measured by inertial observers. That is,
\begin{align}
    \rho_{DE}\geq 0,
\end{align}
in this case the energy-density of the geometric field contribution is measured to be positive according to the WEC. Using the analytical solution (\ref{firstorder}) and the equation (\ref{energydensity}) to obtain the following expression for $\rho_{DE}(z;b)$:
\begin{align}
    \rho_{DE}=\frac{3H^2_{0}\Omega_{\Lambda,0}}{8\pi}\left(1+\frac{3H^2_{0}\Omega^2_{\Lambda,0}b}{2H^2(z;b)}-\frac{15H^4_{0}\Omega^3_{\Lambda,0}b^2}{8H^4(z:b)}\right).
\end{align}
The behavior of the energy density is such that while \(H(z)\) decreases towards the present, \(\rho_{DE}\) grows and deviates considerably from the density of \(\Lambda\). Meanwhile, as \(H(z)\) increases for large values of \(z\), \(\rho_{DE}\) eventually reaches a constant value, coinciding with the density of the cosmological constant. In figure (\ref{darkdensity}) the deviating effect of the parameter \(b\) is shown, which occurs for epochs with redshifts close to the present, while for epochs dominated by radiation and cold matter, \(\rho_{DE}\) becomes the density of \(\Lambda\). For \(b>0\), a minimum occurs in the infinite future with \(\rho_{DE}>0\) always, obeying the WEC, while for \(b<0\), a maximum occurs at the de Sitter point keeping energy density positive, and as expected, the dynamics of the energy density is sensitive to the change in the sign of the parameter.

\begin{figure}[h!]
\includegraphics[width=0.5\linewidth]{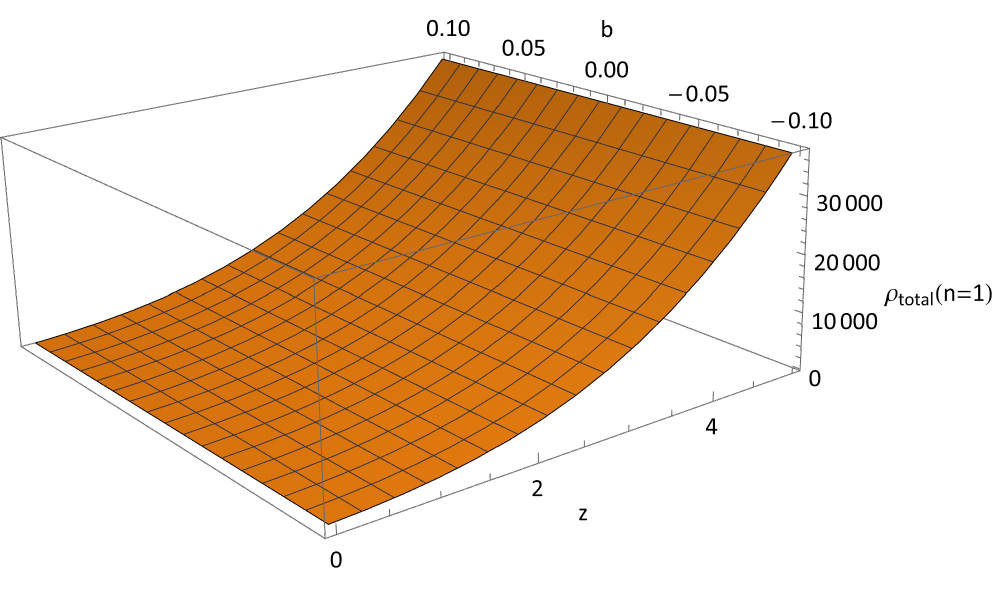} 
\hspace{0.1 cm}
\includegraphics[width=0.5\linewidth]{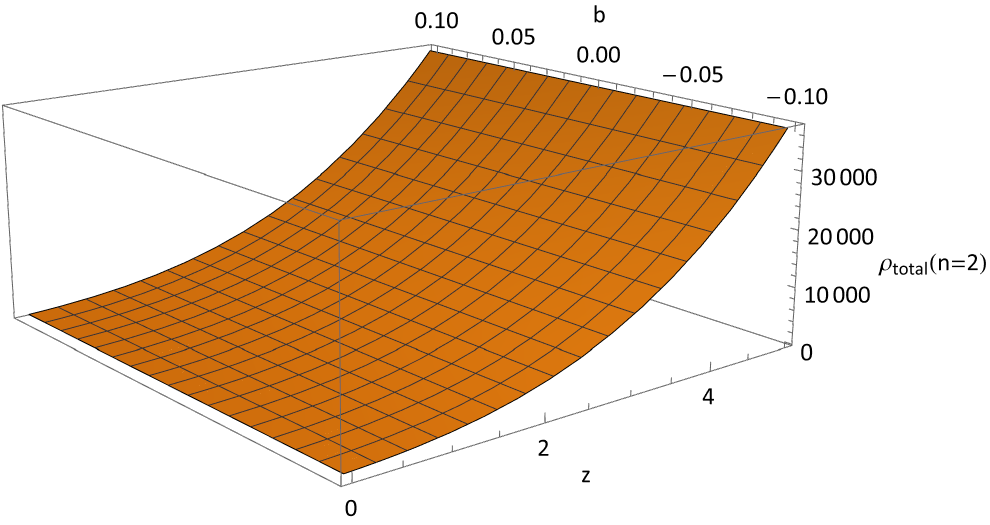}
\caption{Dark energy density $\rho_{DE}$ vs. $(z,b)$ for analytical solution with $n=1$. We have used $\Omega_{m,0}=0.315$ and $H_{0}=67.36 km/s \operatorname{Mpc}^{-1}$ (Left). Dark energy density for analytical solution $n=2$ vs. $(z,b)$ (Right).}
\label{totaldensity}
\end{figure}

For the analytical solution (\ref{secondorder}), the expression for energy-density $\rho^{(n=2)}_{DE}$ is given by:
\begin{align}
\rho^{(n=2)}_{DE}=\frac{3H^2_{0}\Omega_{\Lambda,0}}{8\pi}-\frac{15H^6_{0}\Omega^3_{\Lambda,0}}{32\pi H^4(z;b)}b^2.
\end{align}
We plot this solution with respect to $z$ for positive values of $b$ (the solution is insensitive to the sign) and we show that for this case the WEC is also satisfied, this time with energy-density values lower than that of the cosmological constant (see figure \ref{darkdensity}). What this result shows, has to do with the action $f(Q)$ of the model, which for $n=2$, the exponential $e^{-(H^2_{0}\Omega_{\Lambda,0}b/2H^2(z;b))^2}$ decreases as $H^2(z;b)$ decreases towards infinite future. Therefore the resulting dynamics is an effective cosmological constant with lower energy-density, decreasing its effect towards higher redshifts. In order to determine WEC behaviour for the total contents in the universe, we consider the condition evaluated over the energy-momentum tensor of ordinary matter and the geometrical contribution of the model. Such that,
\begin{align}
T_{\mu\nu}=(\rho_{m}+\rho_{DE}+p_{DE})u_{\mu}u_{\nu}+p_{DE}g_{\mu\nu},
\end{align}
of course we have neglected the contribution from radiation. In order to observe the effect in the evolution with respect to redshift and the parametrization with respect to be $b$ we obtain the plots (\ref{totaldensity}) of the total energy density in the universe with the contribution from the model. The result shows that the WEC is satisfied over $\rho_{total}=\rho_{m}+\rho_{DE}$ for both positive and negative values of $b$. In general, such condition implies that $\rho_{m}+\rho_{DE}\geq 0$, in agreement with the analysis done on the tensor $T^{(DE)}_{\mu\nu}$ it follows that the condition is satisfied for all values of $z$.

\subsection{Strong energy condition (SEC)}
\begin{figure}[h!]
\includegraphics[width=0.5\linewidth]{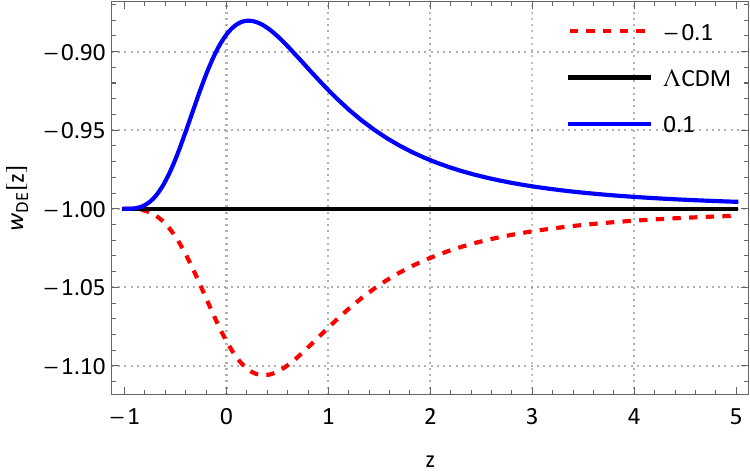} 
\hspace*{0.1 cm}
\includegraphics[width=0.5\linewidth]{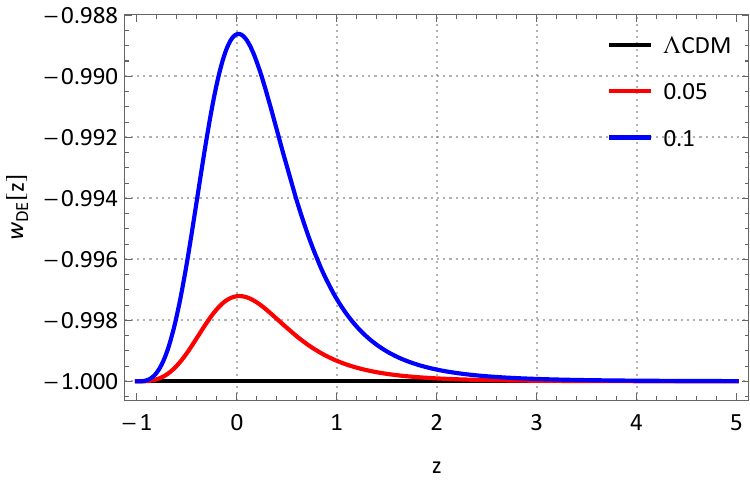}
\caption{Dark energy state parameter vs. $z$ for the solution with $n=1$ for $b=-0.1$ and $b=0.1$ (Left). Dark energy state parameter vs. $z$ for the solution $n=2$ using $b=0.1, 0.05$ (Right)}
\label{stateDE}
\end{figure}

For the SEC, using the general form we obtain:
\begin{align}\label{SECeq}
2\left(T^{(DE)}_{\mu\nu}-\frac{1}{2}T^{(DE)}_{\mu\nu}g_{\mu\nu}\right)u^{\mu}u^{\nu}=\rho_{DE}+3p_{DE}\geq 0,
\end{align}
where it is necessary to compute the dark energy pressure \(p_{DE}\). For this, the relation between \(\rho_{DE}\) and the pressure \(p_{DE}\) via the state parameter \(w_{DE}\) is used:
\begin{align}
w_{DE}=\frac{p_{DE}}{\rho_{DE}}.
\end{align}
This is one of the cosmological parameters that dictates the evolution of the dark energy density, which, up to this point, has a purely geometric origin. It can be calculated using:
\begin{align}
w_{DE}=-1+\frac{(1+z)\rho'_{DE}}{3\rho_{DE}}.
\end{align}
This parameter is $-1$ only when \(\rho_{DE}\) is constant, which occurs for the \(\Lambda\)CDM model. In figure (\ref{stateDE}), the evolution of the state parameter \(w_{DE}\) is shown for different values of \(b\) and both analytical solutions (\ref{firstorder}, \ref{secondorder}), which coincides with the study carried out in \cite{OliverosAcero} for the case \(n=1\). This evolution consists in, for the infinite future, the contribution of the geometry modified by the model asymptotically tends to a cosmological constant in a de Sitter universe for any value of \(b\).

\begin{figure}[h!]
\includegraphics[width=0.5\linewidth]{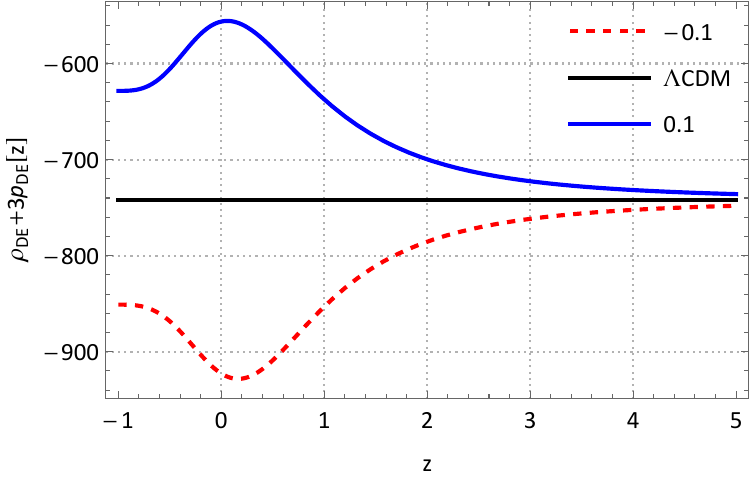} 
\hspace*{0.1 cm}
\includegraphics[width=0.5\linewidth]{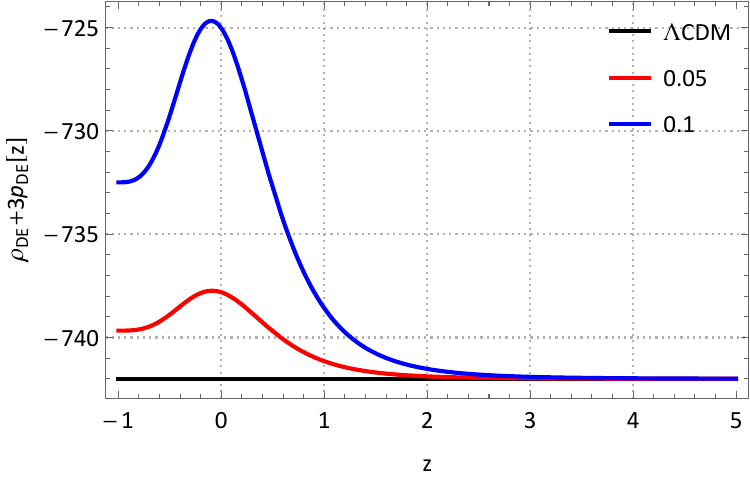}
\caption{SEC for $n=1$, violations are observed for the parameters used (Left). SEC for $n=2$, violations are observed near the $\Lambda$CDM value (Right).}
\label{SECDE}
\end{figure}

\begin{figure}[h]
\includegraphics[width=0.5\linewidth]{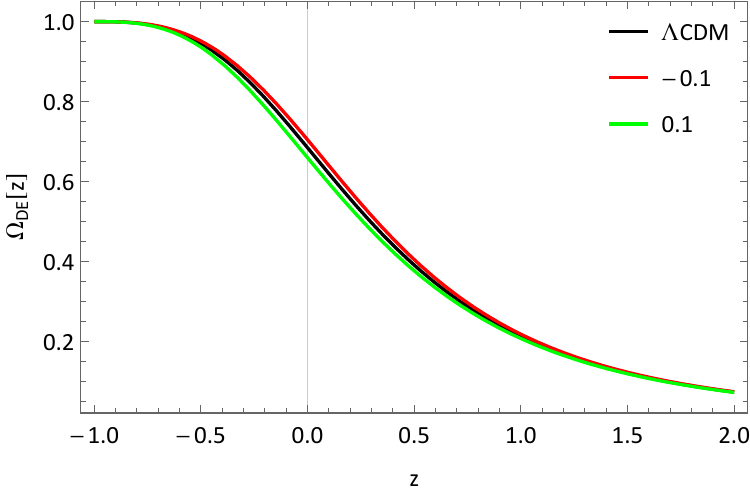} 
\hspace*{0.1 cm}
\includegraphics[width=0.5\linewidth]{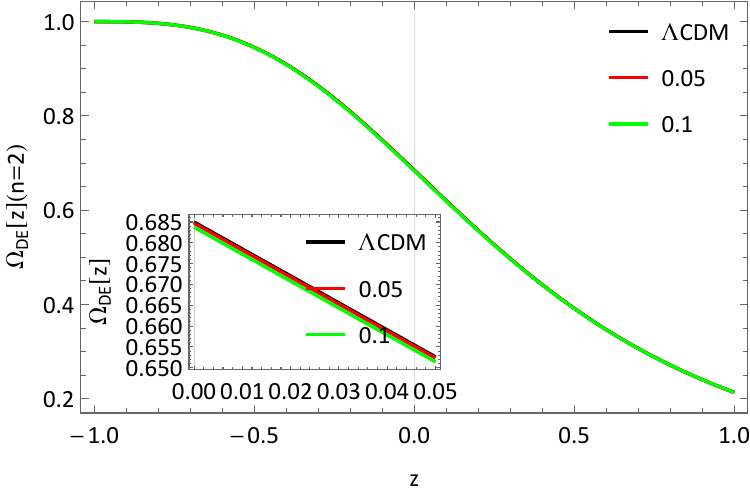}
\caption{$\Omega_{DE}$ for $n=1$, showing greater dominance in the present for $b<0$ than for $b>0$ (Left).  
$\Omega_{DE}$ for $n=2$, where small variations with respect to the $\Lambda$CDM model are observed (Right).}
\label{omegaDE}
\end{figure}

However, near the present, $w_{DE}$ shows variations about $-1$. For $b>0$, we thus have values $w_{DE}>-1$ with the consequence of bounding the pressure from below ($\rho_{DE}>-p_{DE}$) thus the geometrical contribution to dark energy has lesser effect at late times in the expansion of the universe. For $b<0$, the behavior is such that $w_{DE}<-1$, bounding the pressure from above as ($p_{DE}<-\rho_{DE}$) favoring negative values. The contrast between these two behaviors mark that $b>0$ solutions behave as quintessence and $b<0$ as ghost-like fields \cite{copeland, OliverosAcero}. We can ultimately plot the expression (\ref{SECeq}) for the SEC and find the necessary violations for the geometrical contribution alone to produce a regime of accelerated expansion (see figure \ref{SECDE}). We also present the evolution of the $\Omega_{DE}(z)$ parameter, which shows its dominance at late times (asymptotically going to 1), as well as its decreasing value toward the past (for $z\rightarrow \infty$) (\ref{omegaDE}). For $b<0$, we observe that $\Omega_{DE}$ satisfies $\Omega_{DE,0}>\Omega_{\Lambda,0}$, indicating a stronger effect on expansion compared to the cosmological constant alone. Conversely, for $b>0$, the behavior aligns with the analysis carried out about the WEC and $\rho_{DE}$ (see Figure \ref{darkdensity}).

We can also find an analytical expression for $w_{DE}$ up to second order in $b$, for solution $n=2$ and it results as:
\begin{align}
w^{(n=2)}_{DE}(z;b)=-1+\frac{15\Omega^2_{\Lambda,0}\Omega_{m,0}(1+z)^3}{2\xi^{3}(z)}b^2.
\end{align}

At the cosmological present, the state parameter $w^{(n=2)}_{DE}(0;b)$, results in an expression parametrized by $\Omega_{m,0}, \Omega_{\Lambda,0}$. This gives a possible constraint for parameter $b$ as a function of the observable EoS parameter:
\begin{align}
    b=\pm\frac{1}{\Omega_{\Lambda,0}}\sqrt{\frac{2(w_{DE,0}+1)}{15\Omega_{m,0}}}.
\end{align}
In particular, for the general condition $w_{DE}<-1/3$, the parameter is such that $|b|\lesssim 0.775$ using the values as measured by Planck \cite{Planck2018}.

\begin{figure}[h!]
\includegraphics[width=0.5\linewidth]{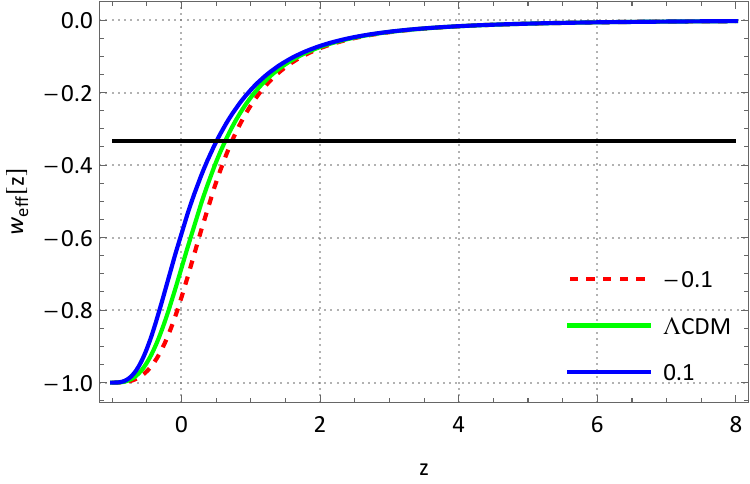} 
\hspace*{0.1 cm}
\includegraphics[width=0.5\linewidth]{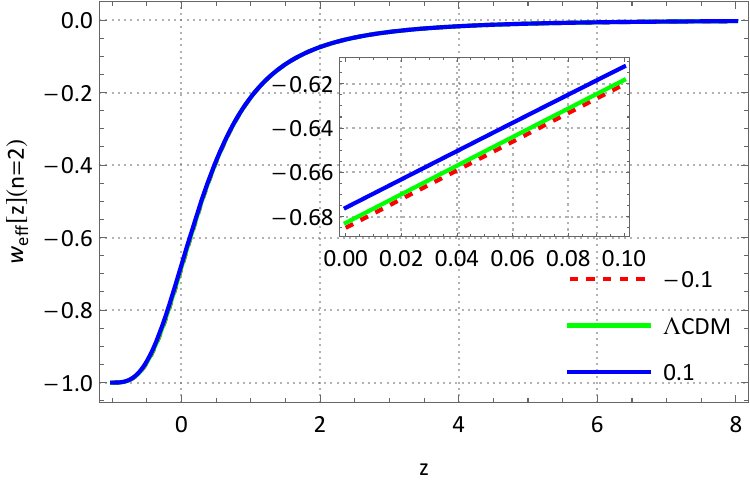}
\caption{$w_{eff}$ for $n=1$, we observe the expected behavior with variations around $\Lambda$CDM at late times and violations to the SEC (Izq). $w_{eff}$ for $n=2$, with a behaviour almost indistinguishable from $\Lambda$CDM. (Der)}
\label{effectivew}
\end{figure}

We use again the SEC general form, but now considering the total energy-momentum tensor $T_{\mu\nu}$, such that $\rho_{m}+\rho_{DE}+3p_{DE}\geq 0$ (with $p_{m}=0$). Therefore we use the effective state parameter $w_{eff}$ to carry out the analysis. This parameter is $-1$ at late times when the geometrical contribution equivalent to dark energy dominates, while asymptotically goes to $0$ at early times with matter domination. This behaviour is shown in the plots (\ref{effectivew}), which using both analytical solutions (\ref{firstorder}, \ref{secondorder}) we obtain violations to the SEC, such that $w_{eff}<-1/3$ is represented by the solid barrier. 

\begin{figure}[h!]
\includegraphics[width=0.5\linewidth]{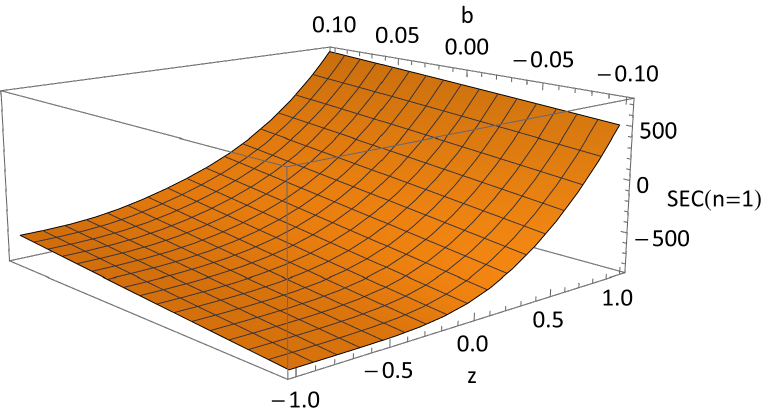} 
\hspace*{0.1 cm}
\includegraphics[width=0.5\linewidth]{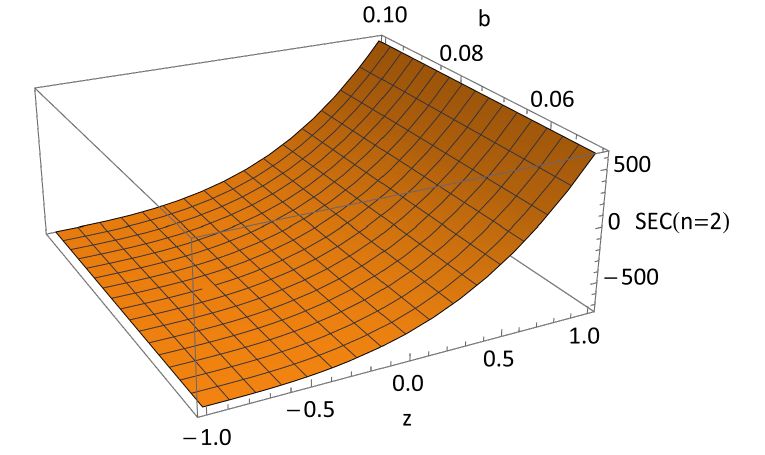}
\caption{Strong energy condition for solution with $n=1$, parametrized by $b$.  (Left). Strong energy condition for $n=2$, parametrized by $b$ (Right.)}
\label{sec3d}
\end{figure}

Similarly, we directly analyze the expression $(T_{\mu\nu}-Tg_{\mu\nu}/2)u^{\mu}u^{\nu}\geq 0$ by observing that violations occurr at late times and near the present. We plot from $z=-1$ to $z=1$ and we obtain clear violations to the SEC for both analytical solutions (\ref{firstorder}, \ref{secondorder}) as shown in figure (\ref{sec3d}). Hence we can see that the geometrical contribution from the model plus in the presence of ordinary matter content induce a cosmological evolution that shows a regime of accelerated expansion at late times, making this in principle, a viable model.

Finally we obtain an analytical expression for $w_{eff}$ in the case of $n=2, b<1$. After several expansions and algebraic manipulations we get,
\begin{align}
    w_{eff}=-1+(1+z)^3\frac{\Omega_{m,0}}{\xi(z)}+\frac{5\Omega^{3}_{\Lambda,0}\Omega_{m,0}(1+z)^3}{2\xi^4}b^2.
\end{align}
Taking the limits at late times, $z\rightarrow -1$ we obtain $w_{eff}\rightarrow -1$ and the limit $z\rightarrow \infty$ we get, $w_{eff}\rightarrow 0$, as expected. Aditionally, at the cosmological present the following expression holds:
\begin{align*}
w_{eff,0}(b)=-1+\Omega_{m,0}+\frac{5\Omega^3_{\Lambda,0}\Omega_{m,0}}{2}b^2.
\end{align*}

\subsubsection{Raychaudhuri's cosmological equation}
\begin{figure}[h!]
\includegraphics[width=0.5\linewidth]{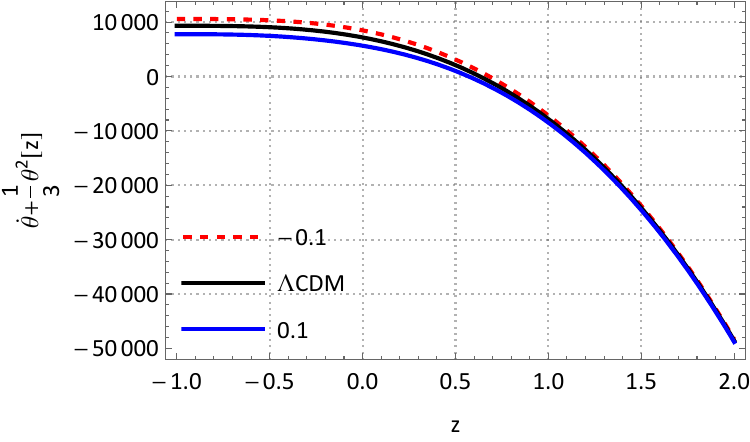}
\hspace*{0.1 cm}
\includegraphics[width=0.5\linewidth]{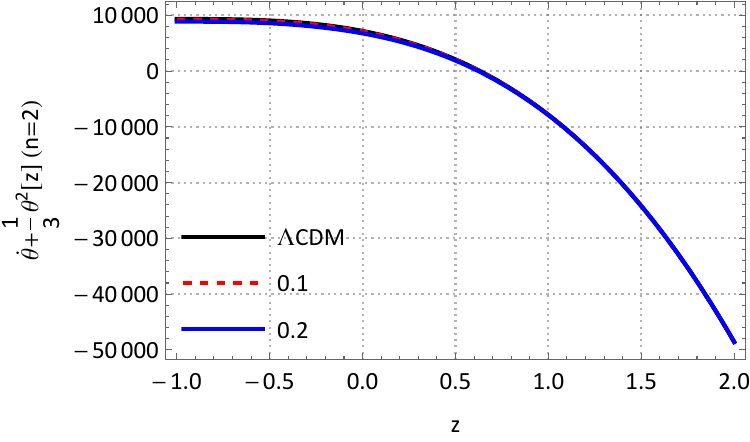} 
\caption{Left-hand side of the Raychaudhuri equation vs. $z$. Using $H(z)$ given by solutions \ref{firstorder}, \ref{secondorder}.}
\label{raychau}
\end{figure}

To complement the study of the strong energy condition, we recall the Raychaudhuri equation for a scalar expansion $\theta(z)=3H(z)$ and calculate $\dot{\theta}$ as follows:
\begin{align}\label{theta}
\dot{\theta}=\dot{z}\theta'=-(1+z)H(z)\theta'\\
\Rightarrow \dot{\theta}=-3(1+z)H(z)H'(z),
\end{align}
The Hubble parameter is parameterized by the set $(b,\Omega_{m,0})$. In general terms, if $H'(z)>0$, the evolution of the scalar expansion tends toward negative values in the past (large values of $z$). In epochs where $H'(z)<0$, the scalar expansion increases the volume of the geodesic congruence in an accelerated manner toward the future. The complete Raychaudhuri equation takes the form:
\begin{align}
-3(1+z)H(z)H'(z)=-3H^2(z)-8\pi \left(T_{\mu\nu}-\frac{1}{2}Tg_{\mu\nu}\right)u^{\mu}u^{\nu}.
\end{align}
This is possible in the coincident gauge \cite{fqreview, trinity}. Grouping the terms with $H(z)$ on the left-hand side and plotting yields figure (\ref{raychau}), where we observe the transition from negative to positive values. For $b<0$, the model deviates prominently from $\Lambda$CDM in the infinite future. 

What do positive and negative values mean? According to the Raychaudhuri equation, the geometrical contribution contributes to positive values, while for the energy-momentum tensor of matter $T^{(matter)}_{\mu\nu}$, the strong energy condition is satisfied and contributes negatively. However, the total energy-momentum tensor is given by:
\begin{align}
T_{\mu\nu}=T^{(matter)}_{\mu\nu}+T^{(DE)}_{\mu\nu},
\end{align}
thus, when separated, the Raychaudhuri equation becomes:
\begin{align}\label{ray}
-3(1+z)H(z)H'(z)+3H^2(z)=-8\pi\rho_{m,0}(1+z)^{3}-8\pi\rho_{DE}(1+3w_{DE}),
\end{align}
which explains that matter dominance produces negative values in the plot. For $b=0$, the geometrical contribution produces an accelerated expansion phase as it begins to dominate, as is well known. However, for $b\neq 0$, the geometric contribution of the model produces earlier positive values for $b=-0.1$, which inevitably leads to the condition $w_{DE}<-1/3$, thus violating the \textbf{SEC}. For $b=0.1$, positive values occur later than in the $\Lambda$CDM model, which aligns with the observed decrease in energy density shown in Figure \ref{darkdensity}. The condition $w_{DE}<-1/3$ indicates that the model used can produce late-time accelerated expansion.

Recalling that $\dot{\theta}+\frac{1}{3}\theta^2=3\ddot{a}/a$, the above expression helps us define the deceleration parameter $q(z)$ as well as other parameters known as \textbf{statefinders} to analyze the late-time cosmological evolution under the exponential model used.

\subsection{NEC and DEC}
Additional to the weak energy condition (WEC) and strong energy condition (SEC), there are other conditions that can be imposed classically over the energy-momentum tensor $T_{\mu\nu}$. A similar case to the WEC occurs when instead of measuring the energy-momentum tensor along timelike curves one considers null or lightlike curves \cite{poisson}.
\\ \\
\textbf{Null energy condition:} The condition is given by the expression $\rho_{m}+\rho_{DE}+p_{DE}\geq 0$ when evaluating the expression $T_{\mu\nu}k^{\mu}k^{\nu}\geq 0$ along a null curve with tangent vector $k^{\mu}$. We can see that in terms of $w_{eff}$, the condition is such that $w_{eff}\geq -1$, which occurs at all previous epochs before the de Sitter point (see figure \ref{effectivew}). Such condition implies a Hubble parameter decreasing in time, for a spatially flat universe. Likewise, from continuity equation it implies, $\dot{\rho}<0$. Generally, scalar fields in nature do not violate such a condition, any cosmology constructed with such fields may produce future singularities and different future scenarios \cite{Rubakov, Beltranull}.
\begin{figure}[h!]
\includegraphics[width=0.5\linewidth]{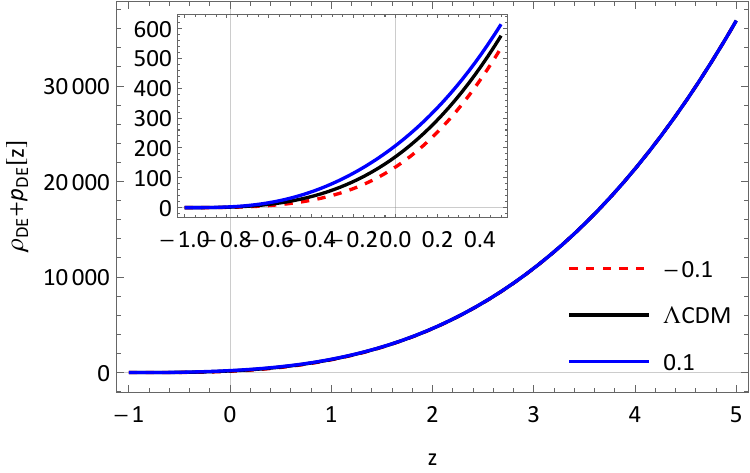}
\hspace*{0.1 cm}
\includegraphics[width=0.5\linewidth]{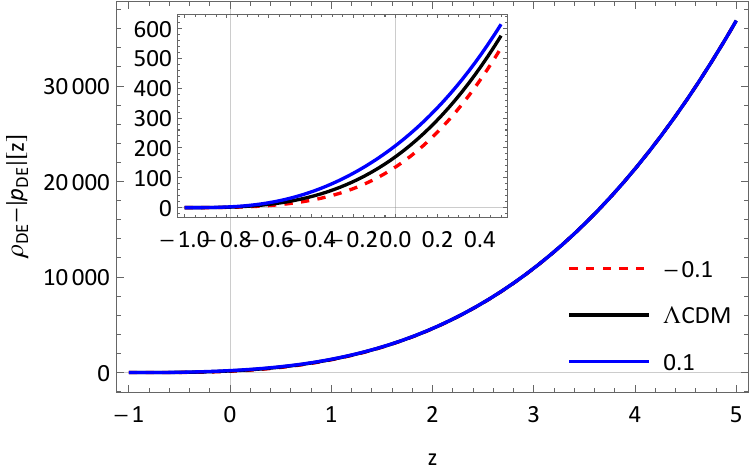} 
\caption{NEC for analytical solution with $n=1$ (Left). DEC for analytical solution with $n=1$, no violations are shown (Right).}
\label{DECNEC}
\end{figure}

\begin{figure}[h!]
\includegraphics[width=0.5\linewidth]{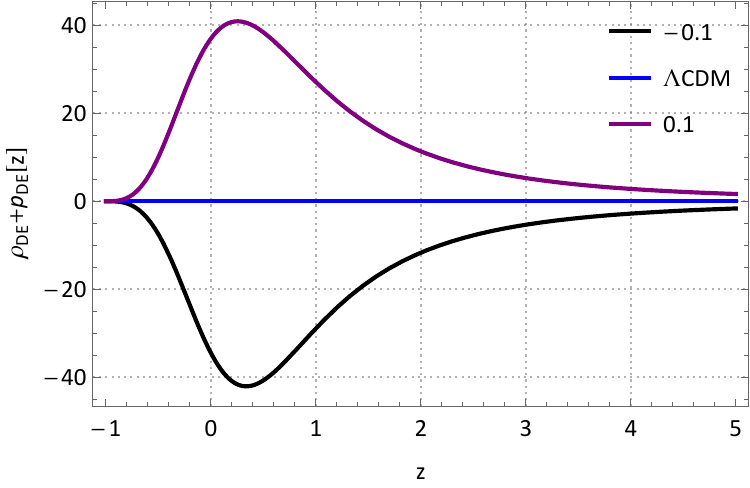}
\hspace*{0.1 cm}
\includegraphics[width=0.5\linewidth]{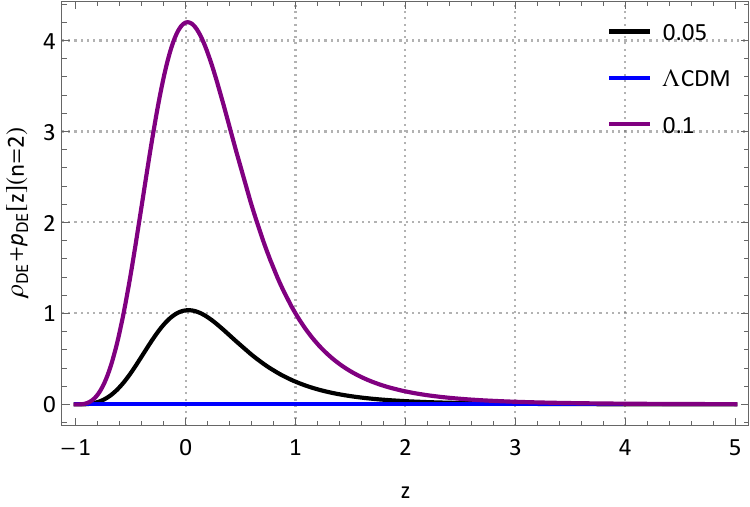} 
\caption{NEC for analytical solution with $n=1$ (Left). NEC for analytical solution with $n=2$ (Der).}
\label{NECDE}
\end{figure}

Additionally, we can study the dark energy component separately, which corresponds to a universe entirely dominated by it. We note that there are violations of the NEC in the analytical solution (\ref{firstorder}) for $b<0$, which confirms the \textit{phantom}-like behavior that could prove problematic for the model in terms of possible instabilities, although such an analysis is beyond the scope of this work. However, for both solutions with $b>0$, the NEC is satisfied (see figure \ref{NECDE}).

\begin{figure}[h!]
\includegraphics[width=0.5\linewidth]{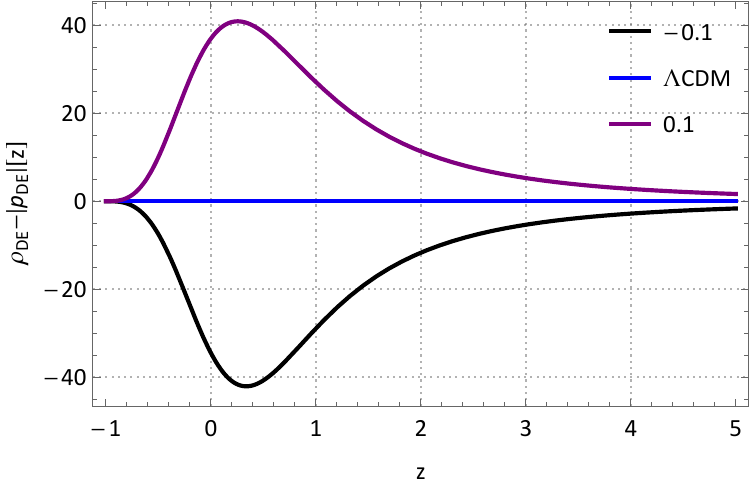}
\hspace*{0.1 cm}
\includegraphics[width=0.5\linewidth]{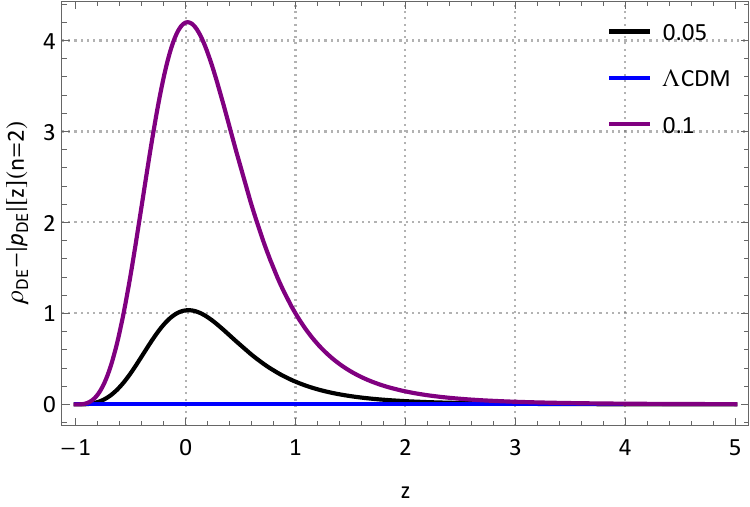} 
\caption{DEC for analytical solution with $n=1$ (Izq). DEC for analytical solution with $n=2$ (Der).}
\label{DECDE}
\end{figure}

\textbf{Dominant energy condition:} While the previous energy conditions are scalar statements, there is one condition that can be imposed in vector form. This is the DEC, which is defined along a \textit{timelike} curve $\gamma'$ with tangent vector $v^{\alpha}$ such that the vector $T_{\alpha\beta}v^{\alpha}$ generally describes the flow of 4-momentum along this curve. Cosmologically, the DEC can be generally expressed as $\rho - |p_{DE}| \geq 0$. From the graph (\ref{DECNEC}), it is shown that this condition is also satisfied for different values of $b$, since, despite the pressure $p_{DE}$ being negative, the absolute value ensures that $\rho \geq |p_{DE}| \geq 0$. We also studied the DEC for the dark energy component separately and found that for $b<0$ in the solution with $n=1$, there are violations consistent with a \textit{phantom}-like field. For $b>0$, the condition is satisfied in both solutions (see figure \ref{DECDE}).

\section{Statefinder parameters}\label{statefinder}

\subsection{Deceleration parameter}
\begin{figure}[h!]
\includegraphics[width=0.5\linewidth]{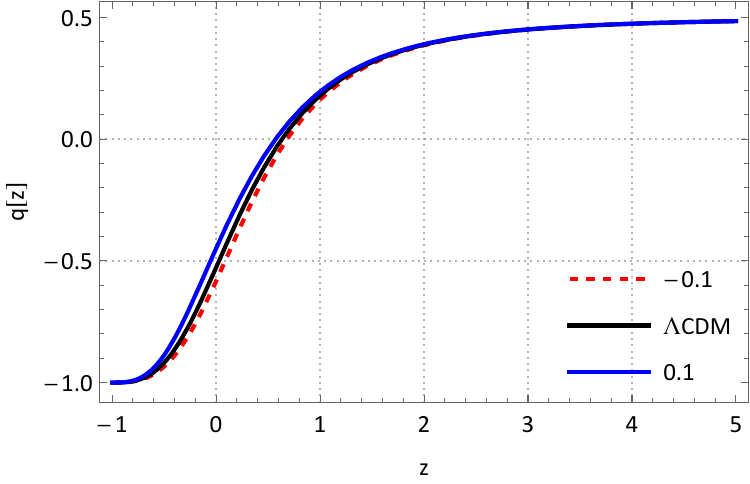}
\hspace{0.1 cm}
\includegraphics[width=0.5\linewidth]{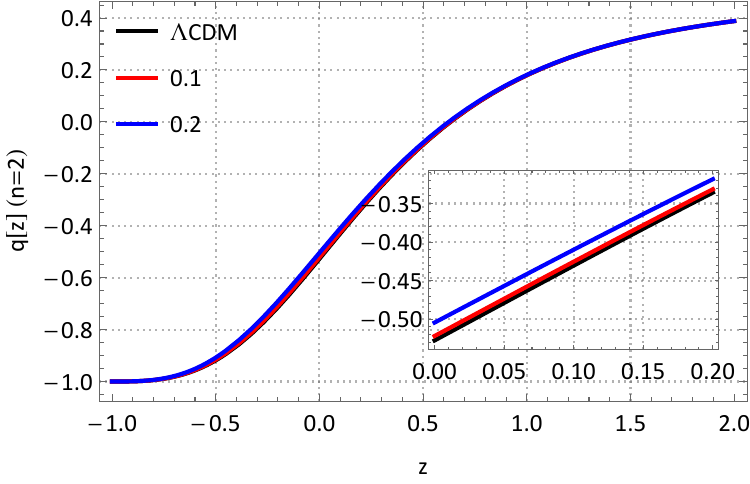} 
\caption{Deceleration parameter $q(z;b)$ vs. $z$. Using $H(z)$ given by solutions \ref{firstorder} [Left], and \ref{secondorder} [Right].}
\label{qparameter}
\end{figure}

The deceleration parameter is defined as:
\begin{align}
q=-\frac{\ddot{a}}{aH^2},
\end{align}
which in terms of $H(z)$, we may obtain $q(z)$ by:
\begin{align}
q(z;b)=-1+(1+z)\frac{H'(z)}{H(z)}.
\end{align}
We expect a transition from positive values (deceleration) to negative values (acceleration), which would indicate that the geometric contribution of the model leads to a late-time accelerated expansion. An analytical expression can be obtained by direct calculation and neglecting terms of higher order than $b^2$, using the analytical solution (\ref{secondorder}) for $H(z)$:
\begin{align}
q(z;b)=-1+\frac{3(1+z)^3\Omega_{m,0}(1+5\Omega^3_{\Lambda,0}b^2/2\xi^3)}{2(\xi-5\Omega^3_{\Lambda,0}b^2/4\xi^2)}\\
\nonumber
\approx -1+\frac{3(1+z)^3\Omega_{m,0}(1+5\Omega^3_{\Lambda,0}b^2/2\xi^3)}{2\xi}\left(1+\frac{5\Omega^3_{\Lambda,0}b^2}{4\xi^3}\right). \\
\approx -1+\frac{3(1+z)^3\Omega_{m,0}}{2\xi(z)}+\frac{45(1+z)^3\Omega^{3}_{\Lambda,0}\Omega_{m,0}b^2}{8\xi^4(z)}.
\end{align}
which is dimensionless and asymptotically tends to $q(\infty)=1/2$ at the infinite past, while at the infinite future it tends to $q(-1)=-1$, passing through the present $z=0$ at $q(0;b)=-0.573+0.569b^2$. With $b<1$, it is ensured that the model can reproduce an acceleration regime in the present. This value is model-dependent, and according to observations, a value of $q_{0}=-0.51\pm 0.024$ is reported in the $\Lambda$CDM model \cite{Brout}, based on the Pantheon+ analysis using Type Ia supernovae. In the present model, for $b>0$, values occur above $\Lambda$CDM, and for $b<0$, values occur below, as shown in Figure \ref{qparameter}.

\subsection{Statefinder in the $r-s$ plane}
Additionally, the statefinder parameters $(r,s)$ were introduced by Sahni \textit{et al.} in \cite{statefinder}, which are given by,
\begin{align}
r=\frac{\dddot{a}}{aH^3}, \quad s=\frac{r-1}{3(q-1/2)},
\end{align}
where the parameter $r$ is a term in a Taylor expansion of the scale factor $a(t)$, and $s$ is a linear combination of the other parameters $(q, s)$. For the $\Lambda$CDM model, the statefinder parameters take the values $(1,0)$, respectively, corresponding to a point on the $r-s$ plane.
\begin{figure}[h!]
 \includegraphics[width=0.5\linewidth]{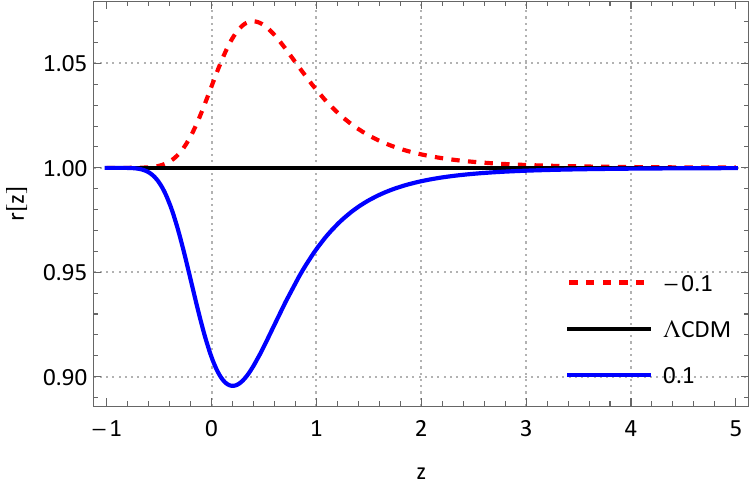}
    \hspace{0.1 cm}
    \includegraphics[width=0.5\linewidth]{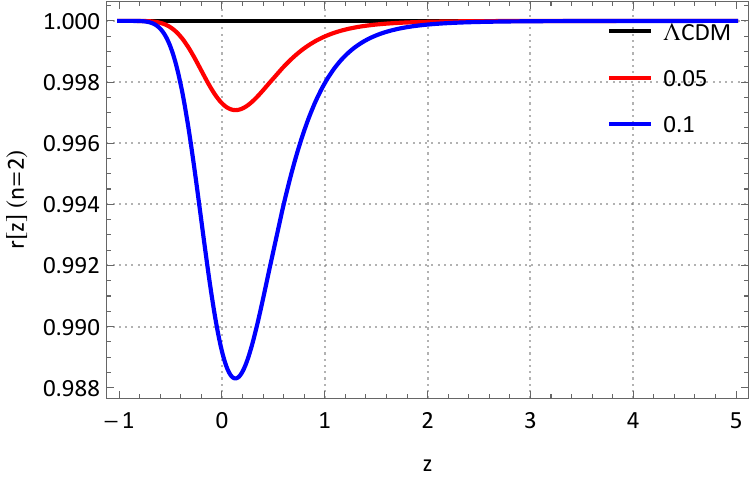}
\caption{Statefinder parameter $r(z;b)$ vs. $z$. For the solutions $n=1$ (Left), and $n=2$ (Right).}
\label{rstatefinder}
\end{figure}

\begin{figure}[h!]
\includegraphics[width=0.5\linewidth]{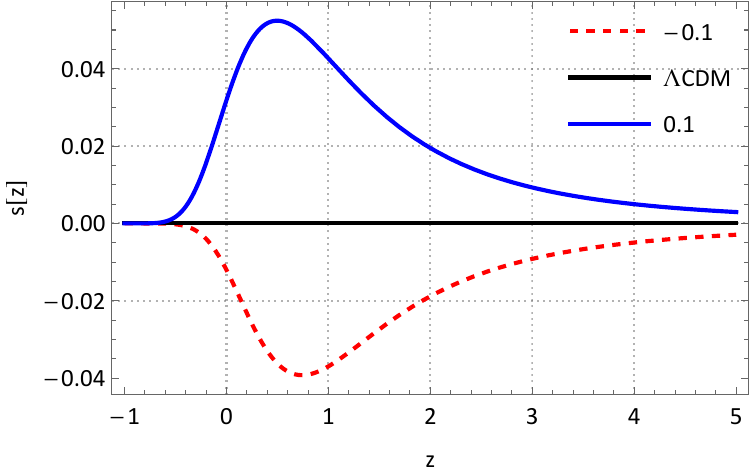}
\hspace*{0.1 cm}
\includegraphics[width=0.5\linewidth]{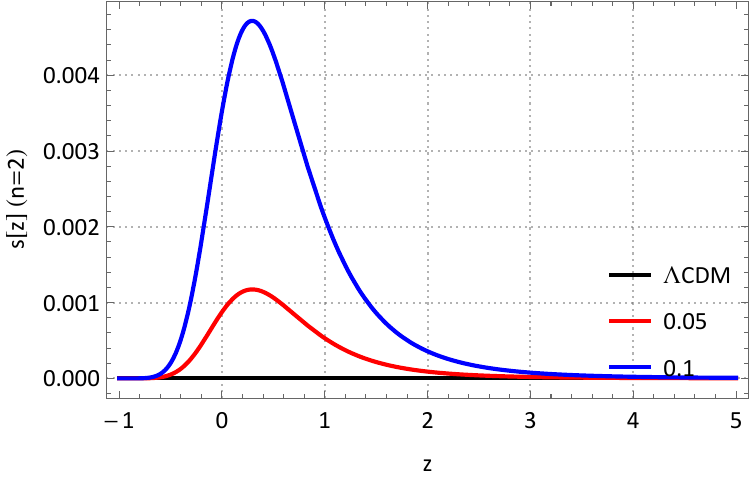} 
\caption{Statefinder parameter $s(z;b)$ vs. $z$. For solutions with $n=1$ [Left],  and $n=2$ [Right].}
\label{statefinders}
\end{figure}

These quantities are geometric as they depend on the scale factor and its derivatives, which allows the study of cosmological evolution through modifications of gravity, as it is the case in our model. In Figures (\ref{rstatefinder}), we find the behavior of the parameters with respect to the redshift $z$. Geometrically, the parameter $r$ can be interpreted through an equivalent expression in terms of the scalar expansion and its derivatives, such that:
\begin{align}
r=1+9\left(\frac{\ddot{\theta}}{\theta^3}+\frac{\dot{\theta}}{\theta^2}\right),
\end{align}
which becomes interesting if we use the focusing theorem from classical gravitation, as for $z\gtrsim 2$, matter begins to dominate over dark energy, driving the scalar expansion to more negative values. Therefore, $\theta<0$, $\dot{\theta}<0$, $\ddot{\theta}<0$, as indicated by the focusing theorem \cite{poisson}. Similarly, $\theta \rightarrow -\infty$ results in the limit $r\rightarrow 1$. A similar situation occurs at late times where $\theta>0$, $\dot{\theta}>0$, $\ddot{\theta}>0$, producing a divergence in the geodesic congruence represented by observers moving with the Hubble flow in the FLRW metric, and similarly, as $\theta\rightarrow +\infty$, the parameter $r$ tends back to $+1$. Note that for the $\Lambda$CDM model, the condition $\ddot{\theta}=-\theta\dot{\theta}$ must be satisfied, that is,
\begin{align}
d_{t}\left(\dot{\theta}+\frac{1}{2}\theta^2\right)=0; \quad \dot{\theta}+\frac{1}{2}\theta^2=C,
\end{align}
this is fulfilled because for Raychaudhuri's equation we have $\dot{\theta}=-9(1+z)^3\Omega_{m,0}H^2_{0}/2$ and $\theta^2=9H^2_{0}[(1+z)^3\Omega_{m,0}+\Omega_{\Lambda}]$ for the $\Lambda$CDM model, so:
\begin{align}
\dot{\theta}+\frac{1}{2}\theta^2=\frac{9H^2_{0}}{2}\Omega_{\Lambda}=C.
\end{align}
It is therefore interesting to note that the dynamics of $r(z)$ are encapsulated in the model's contributions, which for the analytical solution with $n=2$ yields,
\begin{align}
\dot{\theta}+\frac{1}{2}\theta^2=\frac{9H^2_{0}\Omega_{\Lambda,0}}{2}-\frac{45\Omega^3_{\Lambda,0}}{4\xi^3}H^2_{0}(1+z)^3\Omega_{m,0}b^2+\frac{45\Omega^3_{\Lambda,0}}{8\xi^2}H^2_{0}b^2.
\end{align}
Clearly, it is not a constant. In the case of the statefinder parameter $s(z;b)$ of the model, we obtain the behaviors shown in Figures \ref{statefinders}, indicating that for the solution with $n=1$, there is a sensitivity to the sign of the parameter $b$. Comparing the values of the pair $(r,s)$, we find that the model behaves like a \textit{phantom}-type field for negative values of $b$, as $r>1$, $s<0$, and exhibits a \textit{quintessence}-type behavior for positive values of $b$

\begin{figure}[h!]
\centering
\includegraphics[width=0.4\linewidth]{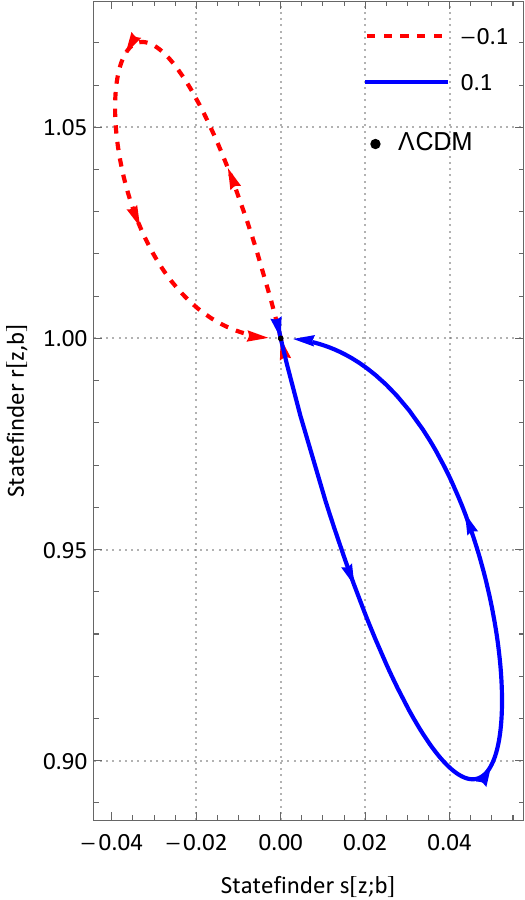}
\hspace{0.1 cm}
\includegraphics[width=0.35\linewidth]{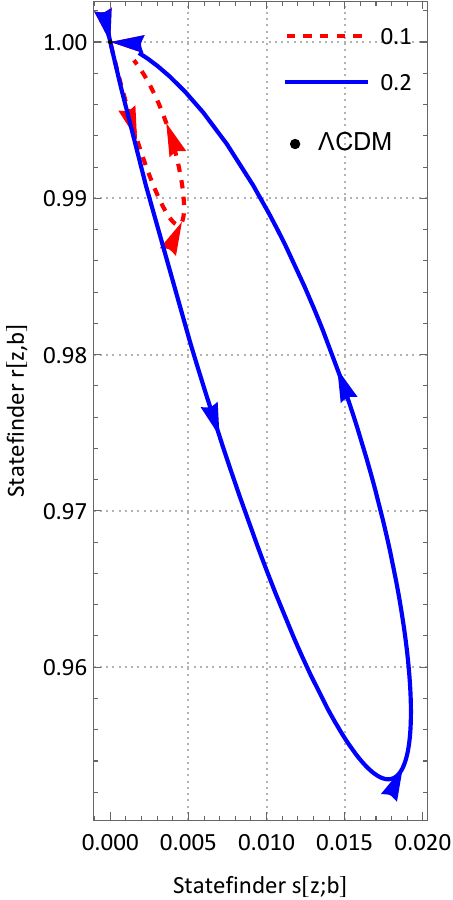} 
\caption{Statefinder trajectories $r(z;b)$ vs. $s(z;b)$. For solutions with $n=1$ [Left],  and $n=2$ [Right].}
\label{rsplane}
\end{figure}

Also, the statefinder diagnostic can be done by plotting the parameters $(r,s)$ on the $r-s$ plane that shows the evolution trajectory of the model with respect to $\Lambda$CDM, as shown in (\ref{rsplane}). 

As a fixed point of the system we have the $\Lambda$CDM model, and for the $n=1$ solution the behavior starts being the cosmological constant, it shows a dynamical regime in near-present epoch and at late-times, going back to the cosmological constant behaviour again. In the case of $b>0$, this geometrical contribution behaves as a quintessence field as seen before. While for the $b<0$ case the dynamics occur in the region of \textit{phantom} and \textit{Chaplygin gas}. Finally, the $n=2$ solution shows a quintessence behavior only, barely departing from the $\Lambda$CDM fixed point. Similar behavior using several models in the $f(Q)$ scenario has been found in \cite{Solanki}

\subsection{$Om(z)$ diagnostic}
\begin{figure}[h!]
\includegraphics[width=0.5\linewidth]{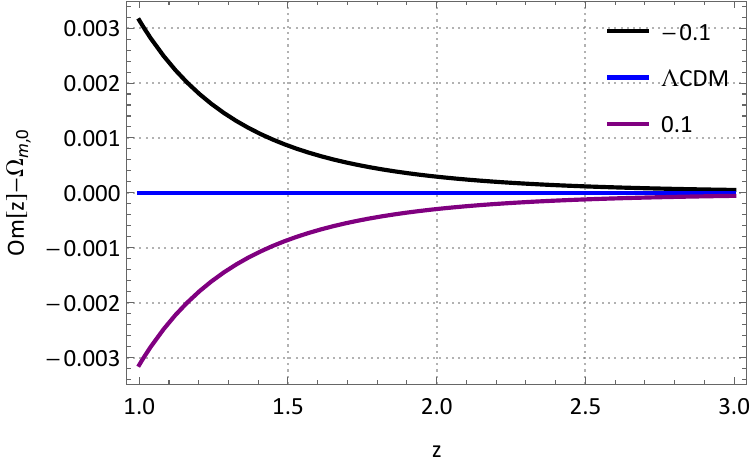}
\hspace*{0.1 cm}
\includegraphics[width=0.5\linewidth]{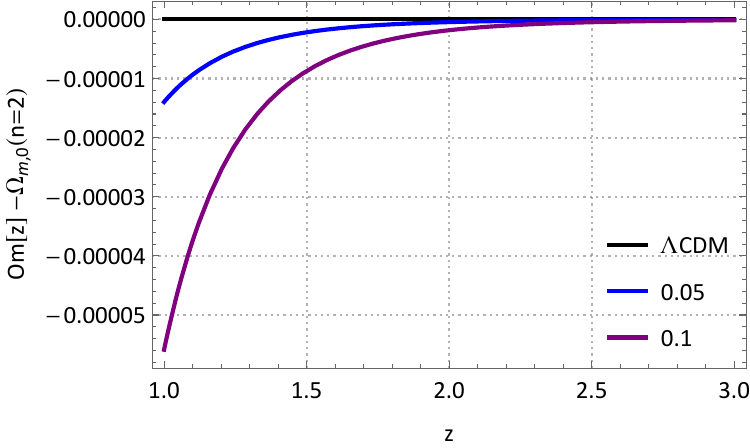} 
\caption{$Om(z)$ Diagnostic. For analytical solutions with $n=1$ [Left],  and $n=2$ [Right].}
\label{omdiagnostic}
\end{figure}

Another interesting diagnostic to characterize the behavior of dark energy in the current model was introduced by Sahni, Starobinsky, and Shafieloo \cite{Sahni2}. Known as the $Om(z)$ diagnostic, it is related to the parameter $\Omega_{m,0}$. Using the Hubble parameter in the $\Lambda$CDM model,
\begin{align}
H^2(z)=H^2_{0}[(1+z)^3\Omega_{m,0}+\Omega_{\Lambda,0}],
\end{align}
where other contributions, such as radiation, are ignored at late times. Considering the relation $\Omega_{\Lambda,0}=1-\Omega_{m,0}$, we have:
\begin{align}
\frac{H^2(z)}{H^2_{0}}-1=\Omega_{m,0}[(1+z)^3-1].
\end{align}
Thus, the quantity $Om(z)$ in the $\Lambda$CDM model is defined as:
\begin{align}
Om(z)=\frac{E^2(z)-1}{(1+z)^3-1}=\Omega_{m,0}; \quad E(z)=H(z)/H_{0}.
\end{align}
This result indicates that any deviation from $\Omega_{m,0}$ signals a different dark energy model, and its nature will depend on the equation of state parameter of that model. In our case, we have found that the dark energy density parameter is given by:
\begin{align}
\Omega_{DE}=\Omega_{DE,0}(1+z)^{3(1-\alpha)},
\end{align}
where $\alpha<1$ for the analytical solution $n=1$ with a positive value of $b$, and $\alpha>1$ for positive values of $b$. Meanwhile, for the solution with $n=2$, $\alpha\gtrsim 1$, with values very close to 1. Therefore, with such a contribution, we have:
\begin{align}
Om(z)=\Omega_{m,0}\frac{(1+z)^3-(1+z)^{3(1-\alpha)}}{(1+z)^3-1}-\frac{(1+z)^{3(1-\alpha)}-1}{(1+z)^3-1}; \quad \Omega_{m,0}+\Omega_{DE,0}=1.
\end{align}
We recover the $\Lambda$CDM model when $\alpha=1$. If the model is \textit{phantom}-like, then the slope of the $Om(z)$ curve is negative, while for quintessence-like models, the slope is positive \cite{Sahni2, Solanki}. Another way to interpret this is to consider that the $\Lambda$CDM model corresponds to the condition $Om(z)-\Omega_{m,0}=0$, where dark energy corresponds to the cosmological constant. In alternative models, $Om(z)-\Omega_{m,0}\neq 0$. We plotted the function $Om(z)-\Omega_{m,0}$ for the mentioned cases and observed (see figure \ref{omdiagnostic}) the behaviors presented by the model. For $b>0$, the curve shows that the geometric contribution corresponds to quintessence-like dark energy, while for $b<0$, it shows a \textit{phantom}-like contribution.

\section{Linear matter perturbations}\label{perturbations}
In this section we complement our study by considering the impact of the model on the growth of linear matter perturbations. As this is a pretty common part in studies exploring modified gravity scenarios, $f(Q)$ is no strange to that. In literature, several signatures beyond the background level have been obtained on cosmological observables \cite{signatures} and, also the impact of a two-parameter model on the large scale structure of the universe by N-body simulations \cite{nbodysimu}. We ought to study key and fundamental features to characterize structure growth of matter perturbations under the previously presented model.

\subsection{Matter perturbations evolution}
\begin{figure}[h!]
\includegraphics[width=0.5\linewidth]{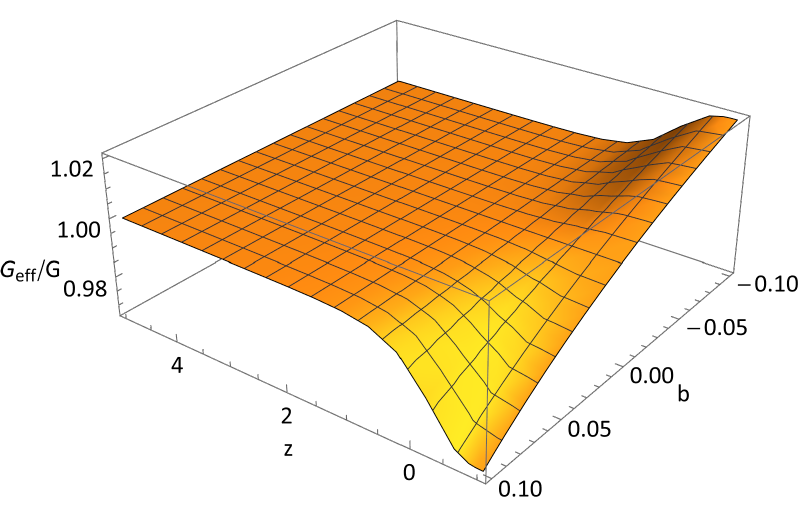}
\hspace*{0.1 cm}
\includegraphics[width=0.5\linewidth]{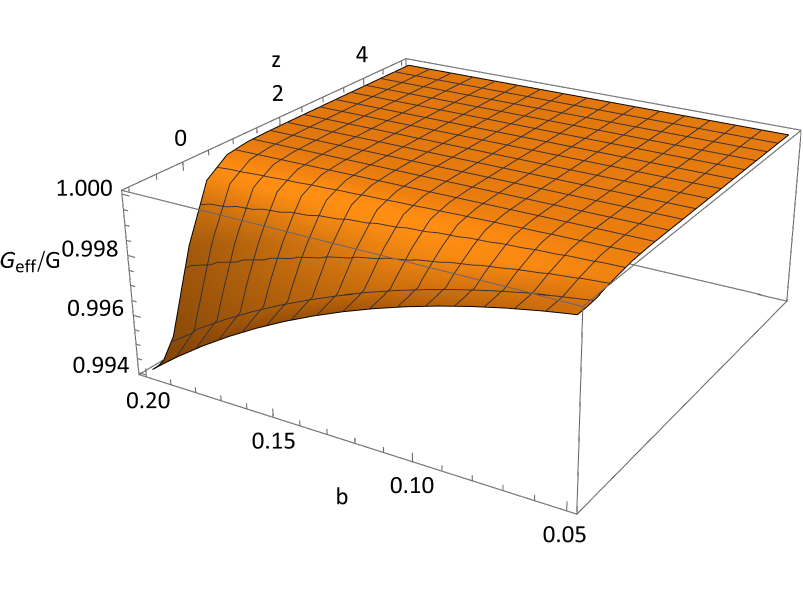} 
\caption{Normalized effective gravitational constant $G_{eff}/G$ vs. $z$. Behavior for $n=1$ [Left] and $n=2$ [Right]. We take some values $b<1$.}
\label{geff}
\end{figure}

\begin{figure}[h!]
\includegraphics[width=0.5\linewidth]{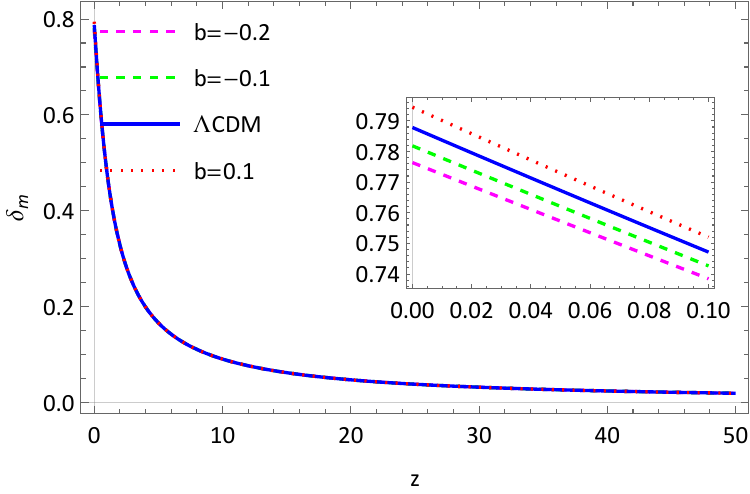}
\hspace*{0.1 cm}
\includegraphics[width=0.5\linewidth]{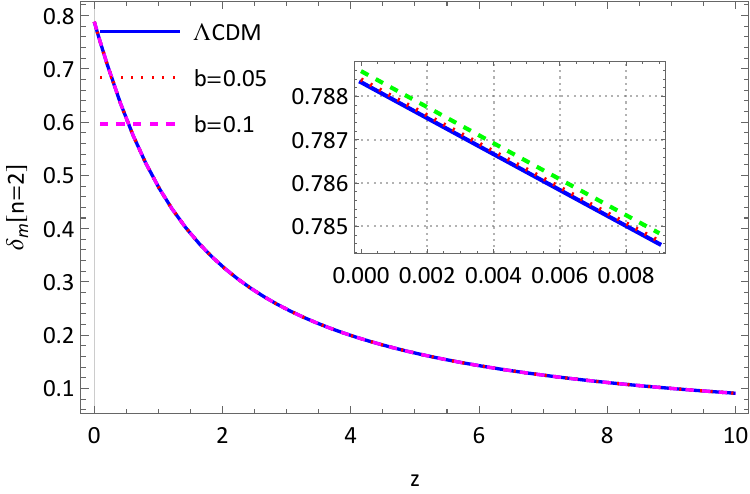}
\caption{Evolución de $\delta_{m}$ vs $z$. Condiciones iniciales $\delta_{m}(50)=\frac{1}{51}$ y $\delta'_{m}(50)=0$ [Izq]. Evolución de las fluctuaciones de densidad $\delta_{m}$ vs $z$ para la solución con $n=2$.[Der]}
\label{linearm}
\end{figure}

Matter perturbations can manifest through fluctuations in local density with respect to the background density in the universe. This quantity is the \textbf{density contrast}, which is gauge-invariant given by:
\begin{align}
    \delta_{m}(x^{\mu})=\frac{\delta \rho_{m}}{\bar{\rho}_{m}},
\end{align}
where $\bar{\rho}_{m}$ is the background density of matter. The resulting function is dependent not only on the evolution parameter (let it be redshift, conformal time, etc) but also on spatial coordinates. By assuming evolution at scales lower than the cosmological horizon, which constitutes the so-called sub-horizon approximation (i.e. $ka(t)\ll H$), the following differential equation has been found to hold for $\delta_{m}$ \cite{cosmologyfq}:
\begin{align}
    \delta''_{m}+\mathcal{H}\delta'_{m}-\frac{4\pi \bar{\rho}G a^2}{d_{Q}f(Q)}\delta_{m}=0,
\end{align}
where we have recovered the gravitational constant $G$ for purpose of the study. This equation contains derivatives with respect to the conformal time, $\delta'_{m}$. In our model, the derivative $d_{Q}f$ results in:
\begin{align}
 d_{Q}f=1+2\Lambda e^{-(b\Lambda/Q)^{n}}(b\Lambda)^{n}(Q)^{-n-1}.
\end{align}
Which for $n=1$ and $n=2$ gives,
\begin{align}
d_{Q}f_{n=1}=1+\frac{H^{4}_{0}\Omega^{2}_{\Lambda,0}b}{4H^4(z)}e^{-H^2_{0}\Omega_{\Lambda,0}b/2H^2}, \\
d_{Q}f_{n=2}=1+\frac{H^{6}_{0}\Omega^{3}_{\Lambda,0}b^2}{8H^{6}(z)}e^{-(bH_{0}\Omega_{\Lambda,0}/2H^2)^{2}}.
\end{align}
It is thus clear that for $b=0$, these derivatives both equal to 1. Otherwise, the evolution is affected by an effective gravitational constant defined as $G_{eff}=G/d_{Q}f$, so that $
G(z)\rightarrow G$ in the $\Lambda$CDM model. An expansion around $b\ll 1$ reveals that,
\begin{align}
d_{Q}f_{n=1}\approx 1+\frac{H^{4}_{0}\Omega^{2}_{\Lambda,0}b}{4H^4(z)}\left(1-\frac{H^2_{0}\Omega_{\Lambda,0}b}{2H^2}\right), \\
d_{Q}f_{n=2}\approx 1+\frac{H^{6}_{0}\Omega^{3}_{\Lambda,0}b^2}{8H^{6}(z)}.
\end{align} 
Which for increasing $H(z)$ in the past, both derivatives leave $G_{eff}\rightarrow G$. On the other hand, near the present the contributions from the model become noticeable. Finally we plot the effective gravitational constant $G_{eff}$ as given by,
\begin{align}
G^{n=1}_{eff}(z;b)=\frac{G}{1+\frac{H^{4}_{0}\Omega^{2}_{\Lambda,0}b}{4H^4(z)}e^{-H^2_{0}\Omega_{\Lambda,0}b/2H^2}}, \\
G^{n=2}_{eff}(z;b)=\frac{G}{1+\frac{H^{6}_{0}\Omega^{3}_{\Lambda,0}b^2}{8H^{6}(z)}e^{-(bH_{0}\Omega_{\Lambda,0}/2H^2)^{2}}}.
\end{align}
We can observe in the plots (\ref{geff}) how the effective constant $G_{eff}$ evolves with respect to the redshift by sweeping values for the parameter $b$. An increasing parameter $b$ decreases the value of $G_{eff}$ for the same redshift in the past, as in this case, the deviation from $\Lambda$CDM becomes progressively smaller. The solutions $n=1, n=2$ behave differently, as for $n=1$, both for positive and negative values of $b$, it holds that $G_{eff} \rightarrow G$ for $z \gtrsim 2$.
However, for $b>0$, the late-time evolution dictates that $G_{eff}<G$, while for $b<0$, it is shown that $G_{eff}>G$ in the future. For $n=2$, $G_{eff}$ rapidly approaches $G$ in the past, while at late times, $G_{eff}\lesssim G$, with values so close that $G_{eff}\sim G$, making it practically indistinguishable from the Newton's gravitational constant. Thus, the influence on the gravitational strength varies effectively along the cosmological evolution, a feature that has an impact on structure formation as it will become evident in following results.
\\ \\
We finally write the evolution differential equation with its derivatives transformed into redshift derivatives by using the chain rule, therefore we get:
\begin{align}\label{linearmeq}
d^{2}_{z}\delta_{m}-[1-(1+z)d_{z}(\ln{H})]\frac{1}{(1+z)}d_{z}\delta_{m}-\frac{3H^2_{0}}{2H^2(z)}\Omega_{m,0}(1+z)G_{eff}(z)\delta_{m}= 0.
\end{align}
The complete evolution at late times is obtained by numerically solving the evolution differential equation for $\delta_{m}(z)$. Here, changes to the $\Lambda$CDM model effectively occur near the present. The same happens for the $n=2$ solution, maintaining the same behavior with the difference of presenting values only above the $\Lambda$CDM model near the present. In both cases, the perturbative nature of the model shows values both above and below $\Lambda$CDM; however, the analytical solution with $n=2$ does not deviate significantly from the $\Lambda$CDM model, with its variation visible only for redshifts on the order of $z\sim 0.001$ (see figure \ref{linearm}).

In the literature, the differential equation (\ref{linearmeq}) is often rewritten in terms of the \textbf{growth factor function}, defined as:
\begin{align*}
f_{g}(z)=\frac{d\ln\delta_{m}}{d\ln a}=-(1+z)\frac{d \ln \delta_{m}}{dz},
\end{align*}
Thus, a first-order nonlinear equation for $f_{g}$ is obtained, as:
\begin{align}
d_{z}f_{g}-\frac{f_{g}}{(1+z)}[2-(1+z)d_{z}(\ln{H})+f_{g}]+\frac{3H^2_{0}}{2H^2}(1+z)^2G_{eff}(z;b)\Omega_{m,0}=0.
\end{align}

\begin{figure}[h]
\includegraphics[width=0.5\linewidth]{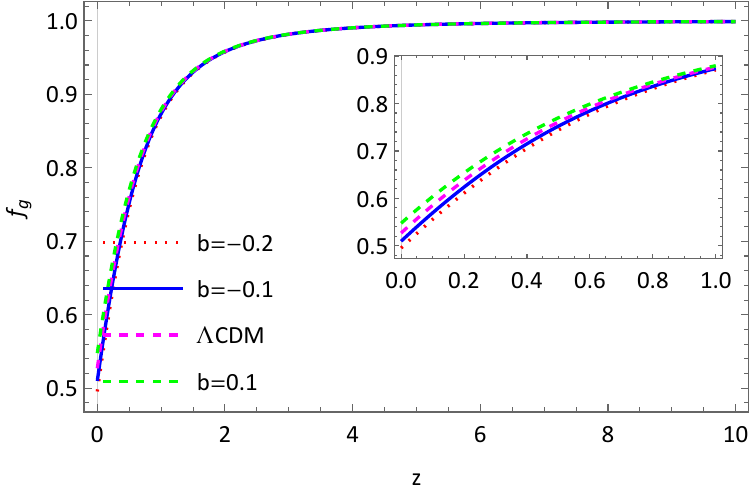}
\hspace{0.1 cm}
\includegraphics[width=0.5\linewidth]{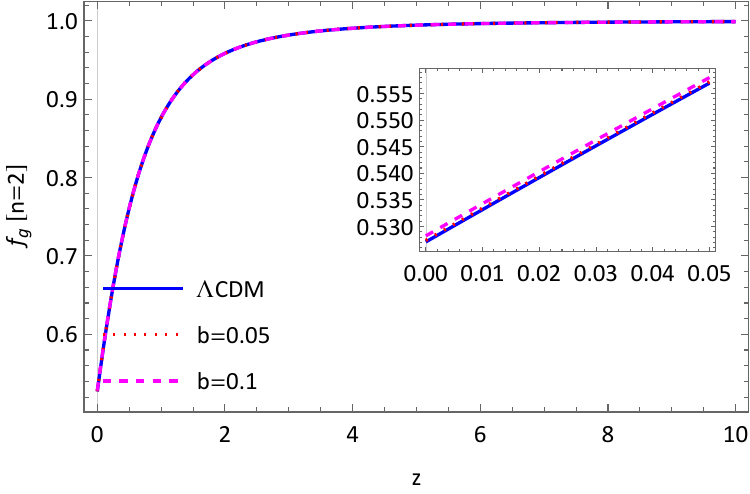}
\caption{Growth factor evolution $f_{g}(z)$ vs $z$. With initial condition $f_{g}(50)=1$ para $n=1$ [Left]. Growth function vs. $z$ para $n=2$ [Right].}
\label{growthfunc}
\end{figure}

We solve this differential equation numerically with the initial condition $f_{g}(50)=1$, obtaining a behavior that asymptotically tends to $f_{g}(z)=1$ during the matter-dominated era, as $\delta_{m}$ evolves as $\delta_{m}\propto a$ (see figure \ref{growthfunc}). Our model is perturbative around $\Lambda$CDM, so it only exhibits appreciable variations at redshifts close to the present, depending on the parameter $b$. 

\begin{figure}[h]
\includegraphics[width=0.5\linewidth]{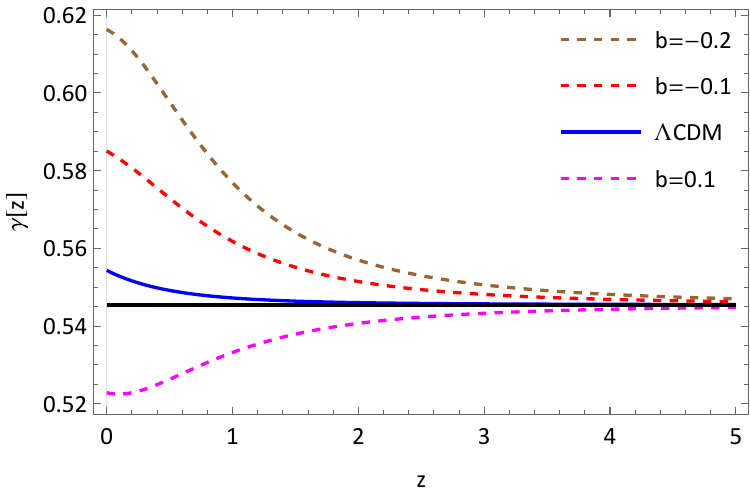}
\hspace*{0.1 cm}
\includegraphics[width=0.5\linewidth]{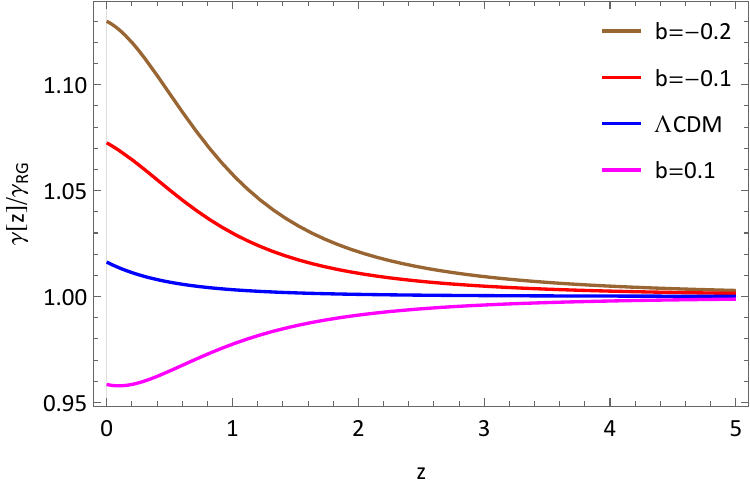}
\caption{Evolution of growth index $\gamma(z)$ vs $z$ for solution with $n=1$. Where the solid black line represents $\gamma=6/11$ as in $\Lambda$CDM [Left] Evolution of normalized growth function $\gamma(z)/\gamma_{GR}(z)$ vs. $z$ for analytical solution $n=1$. Where $\gamma_{RG}$ normalizes the function using $\gamma_{RG}=6/11$ during the matter dominance epoch, here we use $\Omega_{m,0}=0.315$ [Right]}
\label{gammaindex}
\end{figure}

Note that this function characterizes the growth rate of matter perturbations during cosmological evolution. Consequently, it is often reparameterized in such a way that, starting from a known factor such as $\Omega_{m}(z)$, one can determine how $\delta_{m}$ evolves according to the theory of gravity describing the cosmological evolution. Thus, the following form introduced by Wang and Steinhardt \cite{Wang, Khyllep} is often used:
\begin{align}
f_{g}(z)= |\Omega_{m}(z)|^{\gamma(z)}.
\end{align}
For the $\Lambda$CDM model, $\gamma$ is constant and given by $\gamma=6/11$ during the matter-dominated era \cite{Nesseris}. Regarding the model, variations occur at present, as expected, and exhibit the same asymptotic behavior at $\gamma=6/11$ during the matter-dominated era, where the model reduces to $\Lambda$CDM. Once the function $f_{g}(z;b)$ is numerically obtained, the values of $\gamma(z)$ can be derived using the expression $\gamma(z_{i};b)=\ln{f_{g}(z_{i};b)}/\ln{\Omega_{m}(z_{i})}$, yielding the behavior shown in figure \ref{gammaindex}.

Once again, one notices in the plot (\ref{gammaindex}), which was normalized using $\gamma_{RG}=6/11$, how the $f(Q)$ model tends to converge to the value of the $\Lambda$CDM model (from general relativity, GR or equivalently STEGR) in the past. However, near the present, the normalized function is sensitive to the sign change of the parameter $b$. For $b>0$, where the contribution of the model behaves like a quintessence field, the growth index is higher, achieving $\gamma_{f(Q)}(0)>\gamma_{RG}(0)$. This favors the structure growth, indicating a weaker gravity compared to general relativity. On the other hand, in the case of $b<0$, the growth index satisfies $\gamma_{f(Q)}(0)<\gamma_{RG}(0)$, showing a stronger gravity than in general relativity, as it does not favor the growth of matter structures. This behavior is evident in the evolution of the effective universal gravitational constant $G_{eff}(z)$ (see figure \ref{geff}). A similar effect can be observed in other $f(Q)$ models when analyzing the growth index \cite{Khyllep}.

In order to complement our study of the structure growth and formation within the model, we opt to analyze an important parameter in cosmology. Namely, $f\sigma_{8}(z)$, which is formed from a composition between the growth function $f_{g}(z)$ and the amplitude of mass fluctuactions $\sigma_{8}(z)$. The former surges as an statistical variance in matter density distribution in spheres with a radius $R=8 h^{-1} \rm{Mpc}$ \cite{massfluc}. Therefore, in observational and theoretical cosmology, $\sigma_{8}$ represents an important parameter for determining the distribution of galaxies in the universe at large scales. Its current value, measured by the Planck satellite, is $\sigma_{8}(0)=0.8062\pm 0.0057$ \cite{Planck2018}. However, observations at lower redshifts show values different from those measured by Planck using CMB anisotropies. This apparent discrepancy is known as the \textit{$S_{8}$ tension}, quantified through the parameter $S_{8}=\sigma_{8}(\Omega_{m}/0.3)^{1/2}$ \cite{S8tension}. 

The parameter $\sigma_{8}(z)$ can be calculated using the expression employed in [arXiv:2405.07361 [gr-qc]]:
\begin{align}\label{sigma8}
\sigma_{8}(z)=\sigma_{8}(0)\frac{\delta_{m}(z)}{\delta_{m}(0)},
\end{align}
which arises from the relationship between the bias factor, $b$, that linearly relates the variance of galactic density $\sigma_{8, gal}$ and the variance of matter density $\sigma_{8,m}$ \cite{Piattella, Hamilton}. Here $\sigma_{8}(0)=0.8062$ and $\delta_{m}$(0) is the function evaluated at redshift $z=0$.

\begin{figure}[h!]
\includegraphics[width=0.5\linewidth]{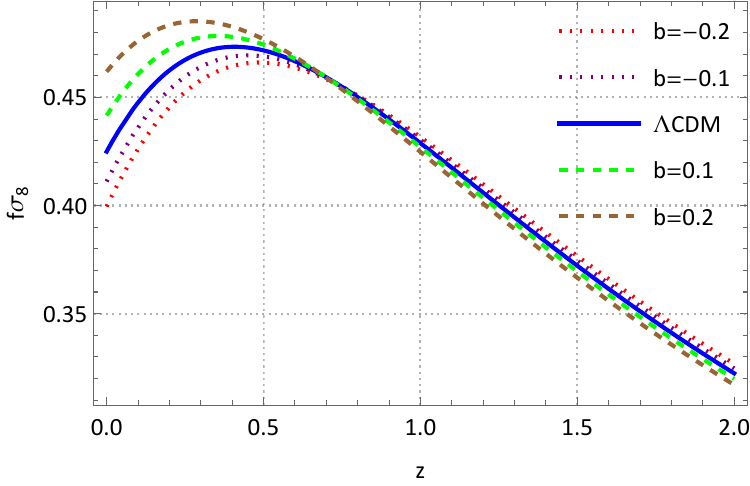}
\hspace*{0.1 cm}
\includegraphics[width=0.5\linewidth]{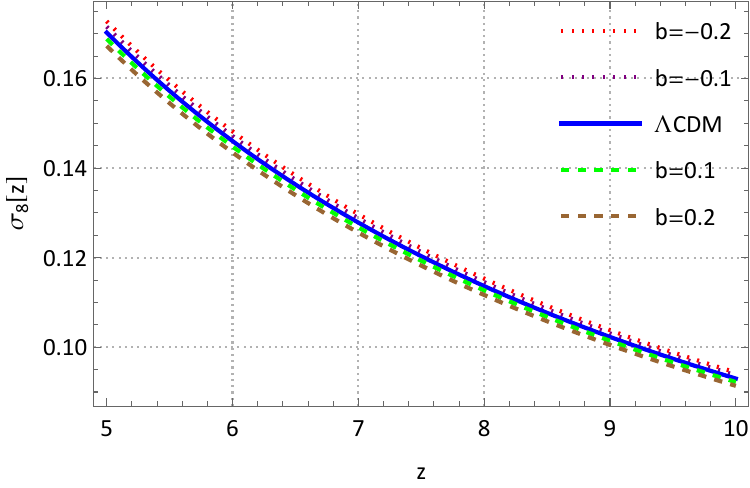}
\caption{$f\sigma_{8}$ vs $z$ evolution up to $z=2.0$ in the case of the analytical solution with $n=1$ [Left].  $f\sigma_{8}(z)$ vs. $z$ evolution from $z=5.0$ to $z=10.0$ in the solution with $n=1$ [Right]}
\label{fsigma8}
\end{figure}

A parameter related to structure growth is the composition of $f(z)$ and $\sigma_{8}(z)$, known as $f\sigma_{8}(z)$. Using both definitions, we obtain:
\begin{align}
f\sigma_{8}(z)=-(1+z)\frac{\sigma_{8}(0)}{\delta_{m}(0)}\frac{d \delta_{m}(z)}{dz}=-(1+z)\frac{d\sigma_{8}}{dz}=\frac{d\sigma_{8}}{d\ln a}.
\end{align}
This parameter thus represents the change in $\sigma_{8}$ with redshift or with the number of e-folds $N=\ln{a}$. We plot $f\sigma_{8}(z)$ for the analytical solution of $H(z)$ with $n=1$ (\ref{firstorder}) (see figure \ref{fsigma8}). Likewise, this parameter is related to redshift distortions, which essentially arise from comparing the position of galaxies in real space and redshift space, where they appear distorted relative to the distribution in real space. At `low redshifts,' the model shows variations near the present and in later times, within the range $z\in (5,10)$, where variations persist throughout the evolution of the parameter. In other words, the evolution of $f\sigma_{8}$ does not asymptotically correspond to the $\Lambda$CDM model and shows that around $\Lambda$CDM, values $b>0$ favor greater matter clustering, while $b<0$ has the opposite effect, dissociating structures. The non-asymptotic behavior of $f\sigma_{8}(z)$ in the model is due to its definition in equation \ref{sigma8}, where $\delta_{m}(z>0)\neq \delta_{m}(0)$ for all positive values of $z$.

\begin{figure}[h!]
\centering
\includegraphics[width=0.7\linewidth]{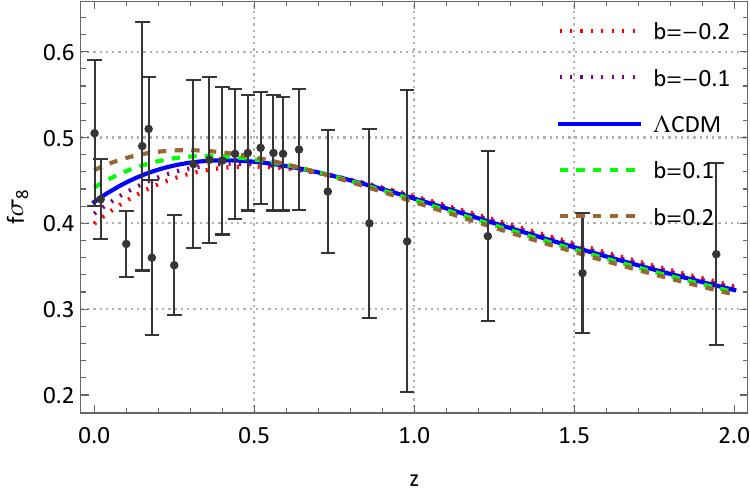}
\caption{Evolution of$f\sigma_{8}$ vs $z$ up to $z=2.0$ in the case of solution $n=1$, in comparison with observational data points obtained from different collaborations and experiments, along with the uncertainty bars.}
\label{fsigma8data}
\end{figure}

To compare the behavior of the model with some observational data of the $f\sigma_{8}$ parameter, we again plot $f\sigma_{8}$ vs. $z$ in comparison with 23 observational data points taken from various references (see \cite{fsigma81, fsigma82, fsigma83, fsigma84, fsigma85, fsigma86, fsigma87, fsigma88, fsigma89, fsigma810, fsigma811,fsigma812}), shown as error bars due to the uncertainty associated with each measurement (Figure \ref{fsigma8data}). This set of observational data comes from various sources, statistically obtained from astrophysical sources such as galaxies, supernovae, among others. The statistical significance as a reliable dataset has been tested in \cite{internal} using Bayesian analysis, as many points considered for comparison are included in the so-called `Gold-2017' dataset obtained in \cite{gold2017} from different surveys. When plotted against the curves calculated from the $f\sigma_{8}(z)$ function for different values of $b$, it can be shown aligns nicely with the observational data, taking into account the associated uncertainty. A key observation here is that we are assigning values for the $b$ parameter, to provide a better fit, a statistical constraint must be performed using this dataset, which has been used in other works \cite{ Barros}, see also \cite{mamdi, sahlu}.

As we mentioned before, the model worked out in this paper have been statistically constrained using $H(z)$ data in \cite{OliverosAcero}, where they performed a statistical analysis using Markov Chain Monte Carlo for the parameter $n=1$. For future works, we would like to perform statistical analysis for the $n=2$ solution using other datasets such as $f\sigma_{8}$.

\section{Conclusions}\label{conclusions}

As of now, the mystery about the existence of dark energy remains unsolved. Although the cosmology community has the $\Lambda$CDM model in their hands, many questions persist. In the mean time, motivated by this mystery, cosmologists keep on proposing new mechanisms to try uncovering the nature of such an entity. Modified gravity scenarios are a plausible solution since they do not require the introduction of new and/or exotic fields, while at the same time they must account for local gravitational phenomena as a correspondence. In that sense, $f(Q)$ theories have acquired some attention in recent years since its introduction in literature and in the same spirit of other modified gravity theories, $f(Q)$ can provide the universe with a regime of accelerated expansion without recurring to additional fields.

In this work we have continued the study of an exponential-type model in $f(Q)$ introduced by one of the authors. In addition to the previous work on the model, in this paper we have extended the analytical solutions for $H(z)$ by considering the parameters $(b\ll 1, n=1)$ and $(b\ll 1, n=2)$. By performing a numerical solution for $H(z)$ the percentage error is such that the analytical solution works well and it can used to analyze cosmological parameters. In this way we have performed analyses on the classical energy conditions as worked out in GR, and found out that for the combination of matter and the geometrical contributions the following holds: (i) WEC is satisfied for both solutions $n=1,2$ and both positive and negative $b$ parameters, (ii) SEC is violated near the present and at late times for both solutions $n=1,2$ producing a regime of accelerated expansion near the present and at late times, (iii) NEC and DEC are satisfied for positive and negative parameters in the solution with $n=1$, showing slight variations very near the present with respect to the $\Lambda$CDM model.  In terms of the isolated energy-momentum tensor $T^{(DE)}_{\mu\nu}$, the following holds: (i) The WEC is satisfied for all values of $b$, (ii) the SEC is violated by positive and negative values of $b$, (iii) the NEC and DEC are violated for $b<0$ and satisfied for $b>0$. In the sense of dark energy models, we find that for $b>0$, a quintessence-like behavior is obtained in the model while for $b<0$, a phantom-like behavior is obtained. 
The dual quintessence-phantom behavior is confirmed when performing a statefinder diagnostic on the model by considering the statefinder quantities $(q,r,s)$ and $Om(z)$. The deceleration parameter readily shows the transition from non-accelerating to an accelerated regime of expansion in the Universe, while an analysis on the $r-s$ plane shows that the trajectories move away from the fixed point $\Lambda$CDM and enter the phantom, quintessence and Chaplygin gas regions in its evolution.

Finally, we complemented our study by including calculations for the linear matter perturbations. In this sense we showed that the effective gravitational constant $G_{eff}(z;b)$ is sensitive to the sign of $b$ for the analytical solution $n=1$, therefore near the present and at late times, this value decreases for $b>0$ and increases for $b<0$ and for the analytical solution $n=2$, it slightly varies from $G$. This proves to have a further impact on quantities as the matter density fluctuations $\delta_{m}$, by giving lower values near the present. Such behavior persists in the growth factor function, for which values $b<0$ give lower values of $f_{g}$ (with respect to $\Lambda$CDM) thus the growth mode evolution for $\delta_{m}$ is slower than just $\delta_{m}\sim a$ and the opposite occurs for $b>0$. Another important quantity in cosmology is $f\sigma_{8}(z)$ which helps in the notion of clustering and structure formation in galaxy clusters, and as we have found, its behavior fits nicely with a dataset of 23 observations made by different collaborations, whose internal robustness has been proven. The fitting does rely on the error bars associated with each measurement, so it is pertaining to build a constraint of the parameter $b$ using such data set and it is a work considered to be done in the future.

After analyzing the model using cosmological parameters at late times, we have shown that it is a viable model that reproduces the main effects of the so-called \textquotedblleft dark energy\textquotedblright. However, $f(Q)$ is recent in the literature and some authors have argued that ghost fields may be propagating due to the presence of additional degrees of freedom in the theory \cite{pathologies1}, but they also found that some theories constructed in symmetric teleparallel geometries (as $f(Q)$) can be constructed in a ghost-free manner \cite{pathologies2}. This, among other arguments against the use of such a theory (see \cite{golovnev}). In that sense, a better understanding of the theory must be gained in the following years to decide further about its use in cosmology and other phenomenological scenarios.


\begin{thebibliography}{99}

\bibitem{supernova1}
A.~G.~Riess \textit{et al.} [Supernova Search Team]: Observational evidence from supernovae for an accelerating universe and a cosmological constant.
Astron. J. \textbf{116}, 1009-1038 (1998)
  
\bibitem{supernova2}
S.~Perlmutter \textit{et al.} [Supernova Cosmology Project]: Measurements of $\Omega$ and $\Lambda$ from 42 High Redshift Supernovae.
Astrophys. J. \textbf{517}, 565-586 (1999)

  
\bibitem{copeland}
E.~J.~Copeland, M.~Sami and S.~Tsujikawa: Dynamics of dark energy.
Int. J. Mod. Phys. D \textbf{15}, 1753-1936 (2006)

\bibitem{Bamba}
K.~Bamba, S.~Capozziello, \textit{et al.}: Dark energy cosmology: the equivalent description via different theoretical models and cosmography tests.
Astrophys. Space Sci. \textbf{342}, 155-228 (2012).

\bibitem{peebles}
P.~J.~E.~Peebles and B.~Ratra: The Cosmological Constant and Dark Energy. Rev. Mod. Phys. \textbf{75} 559-606 (2003)

\bibitem{Einstein1}
A.~Einstein: Cosmological Considerations in the General Theory of Relativity.
Sitzungsber. Preuss. Akad. Wiss. Berlin (Math. Phys. ) \textbf{1917}, 142-152 (1917)

\bibitem{Einstein2}
A.~Einstein and W.~de Sitter: On the Relation between the Expansion and the Mean Density of the Universe.
Proc. Nat. Acad. Sci. \textbf{18}, 213-214 (1932),

\bibitem{lambdaterm}
V.~Sahni and A.~A.~Starobinsky: The Case for a positive cosmological Lambda term.
Int. J. Mod. Phys. D \textbf{9}, 373-444 (2000).

\bibitem{desi}
A.~G.~Adame \textit{et al.} [DESI]: DESI 2024 VI: Cosmological Constraints from the Measurements of Baryon Acoustic Oscillations.
[arXiv:2404.03002 [astro-ph.CO]]. (2024)

  
\bibitem{defelice}
A.~De Felice and S.~Tsujikawa: f(R) theories.
Living Rev. Rel. \textbf{13}, 3 (2010)

\bibitem{Capozziello1}
S.~Capozziello: Curvature quintessence.
Int. J. Mod. Phys. D \textbf{11}, 483-492 (2002)

\bibitem{Clifton}
T.~Clifton, P.~G.~Ferreira,\textit{et al.}: Modified Gravity and Cosmology.
Phys. Rept. \textbf{513}, 1-189 (2012)

\bibitem{fqreview}
L.~Heisenberg: Review on f(Q) gravity.
Phys. Rept. \textbf{1066}, 1-78 (2024)

\bibitem{trinity}
J.~Beltr\'an Jim\'enez, L.~Heisenberg, \textit{et al}: The Geometrical Trinity of Gravity.
Universe \textbf{5}, no.7, 173 (2019)

\bibitem{coincident}
J.~Beltr\'an Jim\'enez, L.~Heisenberg \textit{et al.}: Coincident General Relativity.
Phys. Rev. D \textbf{98}, no.4, 044048 (2018)

\bibitem{hehl}
F.~W.~Hehl, J.~D.~McCrea, \textit{et al.},
:Metric affine gauge theory of gravity: Field equations, Noether identities, world spinors, and breaking of dilation invariance.
Phys. Rept. \textbf{258}, 1-171 (1995)

\bibitem{ortin}
T.~Ortin: Gravity and Strings.
Cambridge University Press (2015)

\bibitem{Myrzakulov}
N.~Myrzakulov, M.~Koussour and D.~J.~Gogoi: A new f(Q) cosmological model with H(z) quadratic expansion.
Phys. Dark Univ. \textbf{42}, 101268. (2023)

\bibitem{MandalQ}
S.~Mandal, S.~Pradhan, \textit{et al.}: Cosmological observational constraints on the power law f(Q) type modified gravity theory.
Eur. Phys. J. C \textbf{83} no.12, 1141 (2023)

\bibitem{anagnos}
F.~K.~Anagnostopoulos, S.~Basilakos \textit{et al.}: First evidence that non-metricity f(Q) gravity could challenge \ensuremath{\Lambda}CDM.
Phys. Lett. B \textbf{822}, 13663  (2021)

\bibitem{anagnos2}
F.~K.~Anagnostopoulos, V.~Gakis, \textit{et al.}: New models and big bang nucleosynthesis constraints in f(Q) gravity. Eur. Phys. J. C \textbf{83} no.1, 58, (2023).

\bibitem{Odintsov1}
S.~Nojiri and S.~D.~Odintsov: Well-defined f(Q) gravity, reconstruction of FLRW spacetime and unification of inflation with dark energy epoch.
Phys. Dark Univ. \textbf{45}, 101538 (2024).

\bibitem{Odintsov2}
S.~Nojiri and S.~D.~Odintsov: F(Q) Gravity with Gauss\textendash{}Bonnet Corrections: From Early-Time Inflation to Late-Time Acceleration.
Fortsch. Phys. \textbf{72} no.9-10, 2400113 (2024).

\bibitem{OliverosAcero}
A.~Oliveros and M.~A.~Acero: Cosmological dynamics and observational constraints on a viable f(Q) nonmetric gravity model. Int. J. Mod. Phys. D \textbf{33}, no.01, 2450004 (2024)

\bibitem{Khyllepexpo}
W.~Khyllep, J.~Dutta, \textit{et al.}: Cosmology in f(Q) gravity: A unified dynamical systems analysis of the background and perturbations. 
Phys. Rev. D \textbf{107} no.4, 044022 (2023).

\bibitem{Cognola}
G.~Cognola, E.~Elizalde, \textit{et al.}: A Class of viable modified f(R) gravities describing inflation and the onset of accelerated expansion.
Phys. Rev. D \textbf{77} (2008), 046009

\bibitem{Odintsov}
S.~D.~Odintsov, D.~S\'aez-Chill\'on G\'omez ,\textit{et al.}: Is exponential gravity a viable description for the whole cosmological history?.
Eur. Phys. J. C \textbf{77} no.12, 862 (2017)

\bibitem{Granda}
L.~N.~Granda: Modified gravity with an exponential function of curvature.
Eur. Phys. J. C \textbf{80}, no.6, 539 (2020)

\bibitem{OliverosAcero2}
A.~Oliveros and M.~A.~Acero: Late-time cosmology in a model of modified gravity with an exponential function of the curvature.
Phys. Dark Univ. \textbf{40}, 101207 (2023)

\bibitem{Oliveros1}
A.~Oliveros: A viable $f (R)$ gravity model without oscillations in the effective dark energy.
Int. J. Mod. Phys. D \textbf{32}, no.12, 2350086 (2023)

\bibitem{Basilakos}
S.~Basilakos, S.~Nesseris and L.~Perivolaropoulos: Observational constraints on viable f(R) parametrizations with geometrical and dynamical probes.
Phys. Rev. D \textbf{87}, no.12, 123529 (2013)

\bibitem{Planck2018}
N.~Aghanim \textit{et al.} [Planck]: Planck 2018 results. VI. Cosmological parameters.
Astron. Astrophys. \textbf{641}, A6 (2020)
[erratum: Astron. Astrophys. \textbf{652}, C4 (2021)]

\bibitem{Riess}
A.~G.~Riess, \textit{et al}: Cluster Cepheids with High Precision Gaia Parallaxes, Low Zero-point Uncertainties, and Hubble Space Telescope Photometry.
Astrophys. J. \textbf{938}, no.1, 36 (2022)

\bibitem{poisson}
  E. Poisson. A relativist's toolkit: the mathematics of black-hole mechanics. Cambridge
University Press, 2004.

\bibitem{Peri}
L.~Perivolaropoulos and F.~Skara: Challenges for \ensuremath{\Lambda}CDM: An update.
New Astron. Rev. \textbf{95}, 101659 (2022)

\bibitem{Mandal}
S.~Mandal, P.~K.~Sahoo and J.~R.~L.~Santos: Energy conditions in $f(Q)$ gravity.
Phys. Rev. D \textbf{102} (2020) no.2, 024057

\bibitem{Capozziello}
S.~Capozziello, S.~Nojiri and S.~D.~Odintsov: The role of energy conditions in $f(R)$ cosmology.
Phys. Lett. B \textbf{781}, 99-106 (2018)

\bibitem{Raychaudhuri}
A.~Raychaudhuri: Relativistic cosmology. 1.
Phys. Rev. \textbf{98} (1955), 1123-1126

\bibitem{Santos}
J.~Santos, J.~S.~Alcaniz, N.~Pires and M.~J.~Reboucas: Energy Conditions and Cosmic Acceleration.
Phys. Rev. D \textbf{75} (2007), 083523

\bibitem{Brout}
D.~Brout, \textit{et al.}: The Pantheon+ Analysis: Cosmological Constraints.
Astrophys. J. \textbf{938} no.2, 110 (2022)

\bibitem{statefinder}
V.~Sahni, T.~D.~Saini, A.~A.~Starobinsky and U.~Alam: Statefinder: A New geometrical diagnostic of dark energy.
JETP Lett. \textbf{77}, 201-206 (2003)

\bibitem{Solanki}
R.~Solanki and P.~K.~Sahoo: Statefinder Analysis of Symmetric Teleparallel Cosmology.
Annalen Phys. \textbf{534} no.6, 2200076 (2022)

\bibitem{Rubakov}
V.~A.~Rubakov: The Null Energy Condition and its violation.
Phys. Usp. \textbf{57}, 128-142 (2014)


\bibitem{Beltranull}
J.~Beltran Jimenez, R.~Lazkoz, \textit{et al.}: Observational constraints on cosmological future singularities.
Eur. Phys. J. C \textbf{76} no.11, 631 (2016)

\bibitem{Sahni2}
V.~Sahni, A.~Shafieloo and A.~A.~Starobinsky: Two new diagnostics of dark energy.
Phys. Rev. D \textbf{78}, 103502 (2008)

\bibitem{signatures}
N.~Frusciante: Signatures of $f(Q)$-gravity in cosmology.
Phys. Rev. D \textbf{103} no.4, 044021 (2021)

\bibitem{nbodysimu}
O.~Sokoliuk, S.~Arora, \textit{et al.}: On the impact of f(Q) gravity on the large scale structure.
Mon. Not. Roy. Astron. Soc. \textbf{522} no.1, 252-267 (2023)

\bibitem{cosmologyfq}
J.~Beltr\'an Jim\'enez, L.~Heisenberg, \textit{et al.}: Cosmology in $f(Q)$ geometry,''
Phys. Rev. D \textbf{101}, no.10, 103507 (2020)

\bibitem{Khyllep}
W.~Khyllep, \textit{et al.}: Cosmological solutions and growth index of matter perturbations in $f(Q)$ gravity.
Phys. Rev. D \textbf{103}, no.10, 103521 (2021) 

\bibitem{Wang}
L.~M.~Wang and P.~J.~Steinhardt: Cluster abundance constraints on quintessence models.
Astrophys. J. \textbf{508}, 483-490 (1998)

\bibitem{Nesseris}
S.~Nesseris and L.~Perivolaropoulos: Testing Lambda CDM with the Growth Function delta(a): Current Constraints.
Phys. Rev. D \textbf{77}, 023504 (2008)

\bibitem{massfluc}
N.~A.~Bahcall and P.~Bode: The Amplitude of mass fluctuations.
Astrophys. J. Lett. \textbf{588}, L1-L4 (2003)

\bibitem{S8tension}
E.~Di Valentino, L.~A.~Anchordoqui \textit{et al.}: Cosmology Intertwined III: $f \sigma_8$ and $S_8$.
Astropart. Phys. \textbf{131}, 102604 (2021)


\bibitem{Piattella}
O.~F.~Piattella: Lecture Notes in Cosmology. Springer (2018).

\bibitem{Hamilton}
Hamilton, A.J.S: Linear Redshift Distortions: A Review. In: Hamilton, D. (eds) The Evolving Universe. Astrophysics and Space Science Library, vol 231. Springer, Dordrecht. (1998)

\bibitem{internal}
B.~Sagredo, S.~Nesseris and D.~Sapone: Internal Robustness of Growth Rate data.
Phys. Rev. D \textbf{98} no.8, 083543 (2018)

\bibitem{gold2017}
S.~Nesseris, G.~Pantazis \textit{et al.}: Tension and constraints on modified gravity parametrizations of $G_{\textrm{eff}}(z)$ from growth rate and Planck data.
Phys. Rev. D \textbf{96} no.2, 023542 (2017)


\bibitem{Barros}
B.~J.~Barros, T.~Barreiro, \textit{et al.}: Testing $F(Q)$ gravity with redshift space distortions.
Phys. Dark Univ. \textbf{30}, 100616 (2020)

\bibitem{mamdi}
D.~Mhamdi, S.~Dahmani, \textit{et al.}: Observational constraints on the growth index parameters in $f(Q)$ gravity.
[arXiv:2408.11996 [gr-qc]] (2024).


\bibitem{sahlu}
S.~Sahlu, \'A.~de la Cruz-Dombriz and A.~Abebe: Structure growth in $f(Q)$ cosmology.
[arXiv:2405.07361 [gr-qc]] (2024).

\bibitem{pathologies1}
D.~A.~Gomes, J.~Beltr\'an Jim\'enez, \textit{et al.}: Pathological Character of Modifications to Coincident General Relativity: Cosmological Strong Coupling and Ghosts in f(Q) Theories.
Phys. Rev. Lett. \textbf{132}, no.14, 141401 (2024)

\bibitem{pathologies2}
A.~G.~Bello-Morales, J.~Beltr\'an Jim\'enez, \textit{et al.}: A class of ghost-free theories in symmetric teleparallel geometry.
JHEP \textbf{12}, 146 (2024)


\bibitem{fsigma81}
C.~Howlett, \textit{et al.}: 2MTF \textendash{} VI. Measuring the velocity power spectrum.
Mon. Not. Roy. Astron. Soc. \textbf{471} no.3, 3135-3151 (2017)

\bibitem{fsigma82}
D.~Huterer, D.~Shafer, \textit{et al.}: Testing $\Lambda$CDM at the lowest redshifts with SN Ia and galaxy velocities.
JCAP \textbf{05}, 015 (2017)

\bibitem{fsigma83}
F.~Shi, X.~Yang, \textit{et al.}: Mapping the Real Space Distributions of Galaxies in SDSS DR7: II. Measuring the growth rate, clustering amplitude of matter and biases of galaxies at redshift $0.1$.
Astrophys. J. \textbf{861} no.2, 137 (2018)

\bibitem{fsigma84}
C.~Howlett, A.~Ross, \textit{et al.}: The clustering of the SDSS main galaxy sample \textendash{} II. Mock galaxy catalogues and a measurement of the growth of structure from redshift space distortions at $z = 0.15$.
Mon. Not. Roy. Astron. Soc. \textbf{449} no.1, 848-866 (2015)

\bibitem{fsigma85}
Y.~S.~Song and W.~J.~Percival: Reconstructing the history of structure formation using Redshift Distortions.
JCAP \textbf{10}, 004 (2009)

\bibitem{fsigma86}
C.~Blake, I.~K.~Baldry, \textit{et al.}: Galaxy And Mass Assembly (GAMA): improved cosmic growth measurements using multiple tracers of large-scale structure.
Mon. Not. Roy. Astron. Soc. \textbf{436}, 3089
(2013)

\bibitem{fsigma87}
L.~Samushia, W.~J.~Percival, \textit{et al.}: Interpreting large-scale redshift-space distortion measurements.
Mon. Not. Roy. Astron. Soc. \textbf{420}, 2102-2119 (2012)

\bibitem{fsigma88}
Y.~Wang, G.~B.~Zhao, \textit{et al.}: The clustering of galaxies in the completed SDSS-III Baryon Oscillation Spectroscopic Survey: a tomographic analysis of structure growth and expansion rate from anisotropic galaxy clustering.
Mon. Not. Roy. Astron. Soc. \textbf{481} no.3, 3160-3166 (2018)

\bibitem{fsigma89}
C.~Blake, S.~Brough, \textit{et al.}: The WiggleZ Dark Energy Survey: Joint measurements of the expansion and growth history at z $\ensuremath{<}$ 1.
Mon. Not. Roy. Astron. Soc. \textbf{425}, 405-414 (2012)

\bibitem{fsigma810}
S.~de la Torre, E.~Jullo, \textit{et al.}: The VIMOS Public Extragalactic Redshift Survey (VIPERS). Gravity test from the combination of redshift-space distortions and galaxy-galaxy lensing at $0.5 < z < 1.2$.
Astron. Astrophys. \textbf{608}, A44 (2017)

\bibitem{fsigma811}
A.~Pezzotta, S.~de la Torre,  \textit{et al.}: The VIMOS Public Extragalactic Redshift Survey (VIPERS): The growth of structure at $0.5 < z < 1.2$ from redshift-space distortions in the clustering of the PDR-2 final sample.
Astron. Astrophys. \textbf{604}, A33 (2017)

\bibitem{fsigma812}
G.~B.~Zhao \textit{et al.} [eBOSS]: The clustering of the SDSS-IV extended Baryon Oscillation Spectroscopic Survey DR14 quasar sample: a tomographic measurement of cosmic structure growth and expansion rate based on optimal redshift weights.
Mon. Not. Roy. Astron. Soc. \textbf{482} no.3, 3497-3513 (2019)
[arXiv:1801.03043 [astro-ph.CO]].

\bibitem{golovnev}
A.~Golovnev: Is there any Trinity of Gravity, to start with?.
[arXiv:2411.14089 [gr-qc]] (2024).

\end{thebibliography}
\end{document}